\documentclass[preprint,12pt]{elsarticle}

\usepackage[margin=3.0cm]{geometry}

\usepackage{graphicx,subfigure}
\usepackage{amssymb}
\usepackage{amsmath}
\usepackage{bm}
\usepackage{lineno}
\usepackage{natbib}
\usepackage[squaren]{SIunits}
\usepackage{color}
\usepackage{soul}
\usepackage{multirow}

\newcommand{\hf}{\frac{1}{2}}
\newcommand{\PP}[2]{\frac{\partial#1}{\partial#2}}
\newcommand{\DD}[2]{\frac{\text{d}#1}{\text{d}#2}}
\newcommand{\pp}[2]{\partial_#2 #1}
\newcommand{\dd}[2]{\text{d}_#2 #1}
\newcommand{\ii}{\mathrm{i}}
\newcommand{\Laplace}{\mathop{}\!\mathbin\bigtriangleup}

\begin{document}

\begin{frontmatter}

\title{A quasi-static particle-in-cell algorithm based on an azimuthal Fourier decomposition for highly efficient simulations of plasma-based acceleration: QPAD }

\author[UCLAEE,SLAC]{Fei Li}
\ead{lifei11@ucla.edu}
\author[BNU]{Weiming An\corref{cor1}}
\ead{anweiming@bnu.edu.cn}
\cortext[cor1]{Corresponding author}
\author[UCLAPH]{Viktor K. Decyk}
\author[SLAC]{Xinlu Xu}
\author[SLAC]{Mark J. Hogan}
\author[UCLAEE,UCLAPH]{Warren B. Mori}

\address[UCLAEE]{Department of Electrical Engineering, University of California Los Angeles, Los Angeles, CA 90095, USA}
\address[UCLAPH]{Department of Physics and Astronomy, University of California Los Angeles, Los Angeles, CA 90095, USA}
\address[SLAC]{SLAC National Accelerator Laboratory, Menlo Park, CA 94025, USA}
\address[BNU]{Department of Astronomy, Beijing Normal University, Beijing 100875, China}


\begin{abstract}
The three-dimensional (3D) quasi-static particle-in-cell (PIC) algorithm is a very efficient method for modeling short-pulse laser or relativistic charged particle beam-plasma interactions. In this algorithm, the plasma response, \emph{i.e.}, plasma wave wake, to a non-evolving laser or particle beam is calculated using a set of Maxwell's equations based on the quasi-static approximate equations that exclude radiation. The plasma fields are then used to advance the laser or beam forward using a large time step. The algorithm is many orders of magnitude faster than a 3D fully explicit relativistic electromagnetic PIC algorithm. It has been shown to be capable to accurately model the evolution of lasers and particle beams in a variety of scenarios. At the same time, an algorithm in which the fields, currents and Maxwell equations are decomposed into azimuthal harmonics has been shown to reduce the complexity of a 3D explicit PIC algorithm to that of a 2D algorithm when the expansion is truncated while maintaining accuracy for problems with near azimuthal symmetry. This hybrid algorithm uses a PIC description in r-z and a gridless description in $\phi$. We describe a novel method that combines the quasi-static and hybrid PIC methods. This algorithm expands the fields, charge and current density into azimuthal harmonics. A set of the quasi-static field equations are derived for each harmonic. The complex amplitudes of the fields are then solved using the finite difference method. The beam and plasma particles are advanced in Cartesian coordinates using the total fields. Details on how this algorithm was implemented using a similar workflow to an existing quasi-static code, QuickPIC, are presented. The new code is called QPAD for QuickPIC with Azimuthal Decomposition. Benchmarks and comparisons between a fully 3D explicit PIC code (OSIRIS), a full 3D quasi-static code (QuickPIC), and the new quasi-static PIC code with azimuthal decomposition (QPAD) are also presented.
\end{abstract}


\end{frontmatter}


\section{Introduction}

Short-pulse, high-intensity laser and beam plasma interaction is an active and robust research area. It involves relativistic, nonlinear and ultrafast plasma physics. It is also a critical topic to the field of plasma based acceleration (PBA). When an intense laser or particle beam propagates through a plasma, it excites a relativistic plasma wave (wakefield). These wakefields support extremely high and coherent accelerating fields which can be more than three orders of magnitude in excess of those in conventional accelerators. The field of PBA has seen rapid experimental progress with many milestones being achieved, including electron acceleration driven by an electron \cite{blumenfeld2007,litos2014}, laser \cite{faure2004,geddes2004,mangles2004,gonsalves2011,wang2013,steinke2016,guenot2017} or proton beam \cite{adli2018}, positron acceleration \cite{corde2015} and PBA-based radiation generation \cite{kneip2010,cipiccia2011,nie2018}.

The rapid progress in experiments has been greatly facilitated by start-to-end simulations using high fidelity particle based methods. The nonlinear aspects of the physics requires the use of fully kinetic tools and the particle-in-cell (PIC) method has proven indispensable. The fully explicit relativistic electromagnetic (EM) PIC method has been used very successfully \cite{dawson1983,birdsall2005,hockney1988}. In this method, individual macro-particles described by Lagrangian coordinates are tracked in continuous phase space as finite size particles (positions and momentum can have continuous values), and then moments of the distribution (\emph{e.g.} charge and current density) are deposited onto  stationary mesh/grid points. The electromagnetic fields are advanced forward in time on the grid points using a discretized version of Maxwell's equations. The new fields are then interpolated to the particles positions to push the particles to new momenta and positions using the relativistic equations of motion. This sequence is repeated for a desired number of time steps.

In most fully explicit PIC codes, a finite-difference time-domain (FDTD) method is used to advance the time-dependent Maxwell's equations and the differential operators are approximated through a finite difference representation (usually accurate to second order of the cell size).  However, to prevent a numerical instability, the time step is constrained by the Courant-Friedrichs-Lewy (CFL) condition which fundamentally couples the spatial and temporal resolution. Roughly speaking, the time step size needs to be less than the smallest cell size which in turn is determined by the smallest physical scale of interest. A second order representation of the time derivative is then used to push the particles. When modeling short-pulse laser and beam-plasma interactions, the moving window technique \cite{decker1994} is always used. In this technique, only a finite window that keeps up with the laser is simulated. New cells and fresh plasma are added to the front, while cells and plasma are dropped off the back. This works because no information and physics that has been dropped can effect the plasma in front of it during the simulation.

Today's supercomputers are capable of providing $\sim10^{16}$ to $\sim10^{17}$ floating point operations per second \cite{top500}. To utilize such computers the algorithm needs data structures that permit many thousands of cores to simultaneously push particles. Effective utilization of such computers has enabled full-scale 3D modeling of intense laser or relativistic charged particles interaction with plasma in some cases. However, even with today's computers and PIC software, it is still not possible to carry out start-to-end simulations of every experiment or proposed concept in full 3D using standard PIC codes. In addition, explicit EM PIC codes can be susceptible to numerical issues including the numerical Cerenkov instability (NCI) \cite{xu2013} and errors to the fields that surround relativistically moving charges \cite{xu2019}. Furthermore, beam loading studies can require very fine resolution in the transverse direction when ion collapse within a particle beam arises \cite{rosenzweig2005,gholizadeh2010,an2017}.

Various methods have been developed to more efficiently model the short-pulse laser and beam-plasma interactions in PBA. These include the boosted frame technique \cite{vay2007}, the quasi-static approximation \cite{mora1997,whittum1997,lotov2003,huang2006,an2013,mehrling2014}, and an azimuthal mode expansion method \cite{lifschitz2009,davidson2015,lehe2016}. The first two are based on the assumption that all relevant waves move forward with velocities near the speed of light, \emph{e.g.}, no radiation propagates backwards. Some of these methods can be combined \cite{yu2016}.

The quasi-static approximation (QSA) was first presented as an analytical tool for studying short-pulse laser interactions \cite{sprangle1990a,sprangle1990b}. The applicability of QSA originates from the disparity in time/length scales between how the laser or particle beam evolves and the period/wavelength of the plasma wake (the plasma response). In the QSA the plasma response is calculated by assuming that the shape of the laser or particle beam (envelope and energy or frequency) is static and the resulting fields from the plasma response are then used to advance the laser or beam forward using a very large time step. It was not until the work of Antonsen and Mora that a PIC algorithm was developed to utilized the QSA. They showed how to push a slice of plasma through a static laser (or move a static laser past a slice of plasma). Their code WAKE \cite{mora1997} is two dimensional (2D) using r-z coordinates and it can model both lasers and particle beams. Whittam also independently developed a QSA PIC code for modeling particle beam-plasma interaction \cite{whittum1997}. In this implementation, it was assumed that plasma particles motion is approximated to be non-relativistic so plasma particles do not move in the beam propagation direction. LCODE \cite{lotov2003} is another 2D r-z PIC code based on the QSA that only models particle beam drivers. QuickPIC \cite{huang2006,an2013} was the first fully 3D QSA based code and it is fully parallelized including a pipelining parallel algorithm \cite{feng2009,an2013}. HiPACE \cite{mehrling2014} is a more recent 3D PIC code based on the QSA. QuickPIC can efficiently simulate both laser pulses and particle beams. It can achieve $10^2$ to $10^4$ speedup without loss of accuracy when compared against fully explicit PIC codes (\emph{e.g.} OSIRIS \cite{fonseca2002}) for relevant problems.

Another recently developed method to enhance the computational efficiency applies the azimuthal Fourier decomposition \cite{lifschitz2009}. In this method, all the field components and current (and charge) density are expanded into a Fourier series in $\phi$ in the azimuthal direction (into azimuthal harmonics denoted by $m$); and the series can be truncated at a value of $m$ determined by the degree of asymmetry for the problem of interest. This algorithm can be viewed as a hybrid method where the PIC algorithm is used in r-z grid and a gridless method is used in $\phi$ and it is sometimes referred to as quasi-3D. By using this algorithm, the problem reduces to solving the complex amplitude (coefficients for Fourier series) for each harmonic on a 2D grid. The complex amplitude, as a function of $r$ and $z$, is updated only at a cost similar to an r-z 2D code. Therefore, if only a few harmonics are kept the algorithm is very efficient. For example, a linearly polarized laser with a symmetric spot size can be described by only the first harmonic. In addition to the much lower cost for advancing fields, much fewer macro-particles are needed for high fidelity. It has been found that speedups of more than two orders of magnitude over a full 3D code are possible.

The quasi-3D method has been implemented into some fully explicit 3D PIC codes \cite{lifschitz2009,davidson2015,lehe2016} and used to study laser \cite{nie2018,ferri2018} and beam \cite{corde2015} plasma interactions. It also been successfully combined with the boosted frame method \cite{yu2016}. However, the azimuthal mode expansion has not been combined with the QSA method or implemented into a quasi-static PIC code. If the quasi-3D technique can be successfully combined with the QSA then dramatic speedups will be possible for problems which are nearly azimuthally symmetric. Such a code will greatly extend the scope of PBA research problems that can be studied numerically.

In this paper, we describe a new code that combines a QSA 3D PIC code with an azimuthal Fourier decomposition, called QPAD (QuickPIC with azimuthal decomposition). The code contains similar procedures and workflow as the 3D quasi-static PIC code QuickPIC, but with the entirely new framework to utilize the azimuthal decomposition. While QuickPIC uses FFT based Poisson solvers to update the fields in each 2D slice of plasma, QPAD computes the fields by means of finite-difference (FD) solvers using the cyclic reduction method \cite{press2010}.
Without loss of accuracy, the code achieves dramatic speedup over fully 3D QuickPIC for a wide range of beam-driven plasma acceleration problems. QPAD currently only supports particle beam drivers.

The paper is organized as follows. In Section \ref{sec:theory}, we derive the governing equations for the complex amplitudes for each harmonic of the relevant fields under the QSA. Section \ref{sec:algorithm} provides details of how the algorithm is implemented. First, the entire numerical workflow that utilizes the three-layer nested loop is described. Next, we introduce the FD implementation of Poisson solvers for each harmonic amplitude and the boundary conditions associated with them. This is followed by a description of the deposition schemes for the source terms for each harmonic needed for the field equations. In Section \ref{sec:simulation}, we compare simulation results between QPAD, QuickPIC and OSIRIS for the beam-driven wakefields and for the hosing instability. A qualitative discussion on the computational speedup is presented in Section \ref{sec:performance}. Lastly, we give a conclusion and a discussion for future work.

\section{Azimuthal decomposition of electromagnetic fields under QSA}
\label{sec:theory}

In this section, we describe the physics arguments behind QPAD including a detailed description of the field equations. As mentioned above, the fundamental differences between a fully explicit 3D PIC code and QPAD are twofold. First, QPAD is a code based on the QSA which separates the time scale of the plasma evolution from that of a drive laser pulse or high-energy particle beam that moves at the speed of light $c$. The assumptions behind the QSA are based on the fact that the characteristic evolution time for a laser driver or a particle beam driver
is several orders of magnitude larger than the plasma oscillation period, $2\pi/\omega_p$ where $\omega_p$ is the plasma frequency. In a quasi-static code, a Galilean spatial transformation is made from $(x,y,z,t)$ (where the laser or beam moves in the $\hat{z}$ direction) to a co-moving frame described by coordinates $(x,y,\xi=ct-z,s=z)$. All the Lagrangian quantities associated with the plasma particles evolve on the fast-varying time-like variable, $\xi$, while those of the beam particles moving close to $c$ evolve on the slow-varying "time" scale, $s$. The transformations $\partial_t=c\partial_\xi,\ \partial_z=\partial_s-\partial_\xi$ are applied for all the Eulerian quantities, \emph{i.e.}, fields, charge density and current density. The QSA assumes that $s$ is the slow-varying time-like scale, \emph{i.e.} $\partial_s\ll\partial_\xi$, so that all the terms associated with $\partial_s$ are small and can thus be neglected.

For remainder of the paper, we use normalized units for all the physical quantities; time, length and mass are normalized to $\omega_p^{-1}$, $c/\omega_p$ and the electron rest mass $m_e$. The normalized Maxwell's equations under the QSA can thus be written as
\begin{align}
\label{eq:maxwell_1}
\nabla_\perp\times \bm{E}_\perp &=-\PP{B_z}{\xi}\bm{e}_z, \\
\label{eq:maxwell_2}
\nabla_\perp\times E_z\bm{e}_z &= -\PP{}{\xi}(\bm{B}_\perp-\bm{e}_z\times\bm{E}_\perp), \\
\label{eq:maxwell_3}
\nabla_\perp\times \bm{B}_\perp - J_z\bm{e}_z &= \PP{E_z}{\xi}\bm{e}_z, \\
\label{eq:maxwell_4}
\nabla_\perp\times B_z\bm{e}_z - \bm{J}_\perp &= \PP{}{\xi}(\bm{E}_\perp+\bm{e}_z\times\bm{B}_\perp), \\
\label{eq:maxwell_5}
\nabla_\perp\cdot\bm{E}_\perp-\rho &= \PP{E_z}{\xi}, \\
\label{eq:maxwell_6}
\nabla_\perp\cdot\bm{B}_\perp &= \PP{B_z}{\xi},
\end{align}
where $\nabla_\perp=\bm{e}_x\partial_x+\bm{e}_y\partial_y$. For convenience, the equations for the transverse and longitudinal fields are written separately. In this context, transverse and longitudinal are defined with respect to the direction of laser or particle beam propagation and not to the direction of the wavenumber of the fields. Taking linear combinations of Eqs. (\ref{eq:maxwell_1}), (\ref{eq:maxwell_3}), (\ref{eq:maxwell_5}) and (\ref{eq:maxwell_6}) leads to equations for the divergence and curl of the transverse force, $\bm{E}_\perp+\bm{e}_z\times\bm{B}_\perp$, on a particle moving at the speed of light along $\hat{z}$,
\begin{align}
\nabla_\perp\times(\bm{E}_\perp+\bm{e}_z\times\bm{B}_\perp)&=\bm{0}, \nonumber \\
\nabla_\perp\cdot(\bm{E}_\perp+\bm{e}_z\times\bm{B}_\perp)&=\rho-J_z. \nonumber
\end{align}
We can infer from the first of these equations  that the transverse force can be described by the transverse gradient of a scalar potential which we call $\psi$,
\begin{equation}
\label{eq:def_psi}
\bm{E}_\perp+\bm{e}_z\times\bm{B}_\perp=-\nabla_\perp\psi.
\end{equation}
Substituting this relationship into the second equation, leads to a Poisson equation for the pseudo potential $\psi$,
\begin{equation}
\label{eq:poisson_psi}
-\nabla_\perp^2\psi=(\rho-J_z).
\end{equation}
By taking $\bm{e}_z \times$ on both sides of Eq. (\ref{eq:maxwell_2}) and using the relation (\ref{eq:def_psi}), it can be inferred that $E_z=\frac{\partial\psi}{\partial\xi}$. This relationship also follows directly from the definition, $E_z=-\frac{\partial \varphi}{\partial z}-\frac{\partial A_z}{\partial t}$, and the QSA, where $\varphi$ and $A_z$ are the scalar potential and the $\hat{z}$-component of the vector potential.

The transverse force $\bm{E}_\perp+\bm{e}_z\times\bm{B}_\perp$ in Eq. (\ref{eq:maxwell_4}) and the quantity $\bm{B}_\perp-\bm{e}_z\times\bm{E}_\perp$ in Eq. (\ref{eq:maxwell_2}) are not independent. Therefore, the quasi-static form of Maxwell's equations given above cannot be used to advance the fields forward in time, \emph{i.e.}, $\xi$, using the FDTD methods as is done in fully explicit PIC codes. Therefore, in QuickPIC, a set of Poisson-like equations are employed to directly solve the fields,
\begin{align}
\label{eq:poisson_bxy}
\nabla_\perp^2\bm{B}_\perp&=\bm{e}_z\times\left(\PP{\bm{J}_\perp}{\xi}+\nabla_\perp J_z\right), \\
\label{eq:poisson_bz}
\nabla_\perp^2 B_z&=-\bm{e}_z\cdot(\nabla_\perp\times\bm{J}_\perp), \\
\label{eq:poisson_ez}
\nabla_\perp^2 E_z&=\nabla_\perp\cdot\bm{J}_\perp.
\end{align}
which can be derived by applying the QSA to the wave equations for $\bm{E}$ and $\bm{B}$. After obtaining $\bm{B}_\perp$ from Eq. (\ref{eq:poisson_bxy}) and $\psi$ from Eq. (\ref{eq:poisson_psi}), we can calculate $\bm{E}_\perp$ by subtracting $\bm{e}_z\times\bm{B}_\perp$ from $-\nabla_\perp\psi$. Although it is not directly used in QuickPIC, for completeness we write out the Poisson-like equation for $\bm{E}_\perp$,
\begin{equation}
\label{eq:poisson_exy}
\nabla_\perp^2\bm{E}_\perp=\nabla_\perp\rho+\PP{\bm{J}_\perp}{\xi}.
\end{equation}

We next expand the electromagnetic fields, charge density and current density in cylindrical coordinates with each quantity being decomposed into a Fourier series in the azimuthal direction. To obtain a set of equations for the Fourier amplitude of each azimuthal harmonic, we first write the field equations, Eqs. (\ref{eq:def_psi})-(\ref{eq:poisson_exy}), in cylindrical coordinates,
\begin{align}
\label{eq:fperp}
\bm{e}_r\PP{\psi}{r}+\bm{e}_\phi\frac{1}{r}\PP{\psi}{\phi}&=(-E_r+B_\phi)\bm{e}_r-(E_\phi+B_r)\bm{e}_\phi, \\
\nabla_\perp^2\psi&=-(\rho-J_z), \\
\label{eq:pois_br}
\nabla_\perp^2B_r-\frac{B_r}{r^2}-\frac{2}{r^2}\PP{B_\phi}{\phi}&=-\PP{J_\phi}{\xi}-\frac{1}{r}\PP{J_z}{\phi}, \\
\nabla_\perp^2B_\phi-\frac{B_\phi}{r^2}+\frac{2}{r^2}\PP{B_r}{\phi}&=\PP{J_r}{\xi}+\PP{J_z}{r}, \\
\nabla_\perp^2B_z&=-\frac{1}{r}\PP{}{r}(rJ_\phi)+\frac{1}{r}\PP{J_r}{\phi}, \\
\nabla_\perp^2E_r-\frac{E_r}{r^2}-\frac{2}{r^2}\PP{E_\phi}{\phi}&=\PP{\rho}{r}+\PP{J_r}{\xi}, \\
\nabla_\perp^2E_\phi-\frac{E_\phi}{r^2}+\frac{2}{r^2}\PP{E_r}{\phi}&=\frac{1}{r}\PP{\rho}{\phi}+\PP{J_\phi}{\xi}, \\
\label{eq:pois_ez}
\nabla_\perp^2E_z&=\frac{1}{r}\PP{}{r}(rJ_r)+\frac{1}{r}\PP{J_\phi}{\phi}.
\end{align}
where the 2D (transverse) Laplacian is defined as $\nabla_\perp^2\equiv\frac{1}{r}\PP{}{r}\left(r\PP{}{r}\right)+\frac{1}{r}\PP{^2}{\phi^2}$. Expanding the electromagnetic fields, charge and current density into a Fourier series in the azimuthal direction, gives
\begin{equation}
\label{eq:fourier_series}
\begin{aligned}
U(r,\phi)&=\sum_{m=-\infty}^{+\infty}U^m(r)e^{\ii m\phi}\\
&=U^0(r)+2\sum_{m=1}\mathfrak{Re}\{U^m\}\cos(m\phi)-2\sum_{m=1}\mathfrak{Im}\{U^m\}\sin(m\phi)
\end{aligned}
\end{equation}
where $U$ represents an arbitrary scalar field or components of a vector field, and note that the amplitude of each harmonic $U^m$ is complex. It follows that $U^{-m}=(U^m)^*$ because $U(r,\phi)$ is real, which indicates that only the evolution of $m\ge0$ modes need to be considered. Substituting the expansion into Eqs. (\ref{eq:fperp})-(\ref{eq:pois_ez}) yields the following governing equations for each mode
\begin{align}
\label{eq:fperp_m}
\bm{e}_r\PP{\psi^m}{r}+\bm{e}_\phi\frac{\ii m}{r}\psi^m&=(-E_r^m+B_\phi^m)\bm{e}_r-(E_\phi^m+B_r^m)\bm{e}_\phi,\\
\label{eq:pois_psi_m}
\Laplace_m \psi^m&=-(\rho^m-J_z^m), \\
\label{eq:pois_br_m}
\Laplace_m B_r^m-\frac{B_r^m}{r^2}-\frac{2\ii m}{r^2}B_\phi^m&=-\PP{J_\phi^m}{\xi}-\frac{\ii m}{r}J_z^m, \\
\label{eq:pois_bphi_m}
\Laplace_m B_\phi^m-\frac{B_\phi^m}{r^2}+\frac{2\ii m}{r^2}B_r^m&=\PP{J_r^m}{\xi}+\PP{J_z^m}{r}, \\
\label{eq:pois_bz_m}
\Laplace_m B_z^m&=-\frac{1}{r}\PP{}{r}(rJ_\phi^m)+\frac{\ii m}{r}J_r^m, \\
\Laplace_m E_r^m-\frac{E_r^m}{r^2}-\frac{2\ii m}{r^2}E_\phi^m&=\PP{\rho^m}{r}+\PP{J_r^m}{\xi}, \\
\Laplace_m E_\phi^m-\frac{E_\phi^m}{r^2}+\frac{2\ii m}{r^2}E_r^m&=\frac{\ii m}{r}\rho^m+\PP{J_\phi^m}{\xi}, \\
\label{eq:pois_ez_m}
\Laplace_m E_z^m&=\frac{1}{r}\PP{}{r}(rJ_r^m)+\frac{\ii m}{r}J_\phi^m
\end{align}
where $\Laplace_m\equiv\frac{1}{r}\PP{}{r}\left(r\PP{}{r}\right)-\frac{m^2}{r^2}$. This set of equations is overdetermined and therefore, similarly to what is currently used in the 3D QuickPIC algorithm \cite{an2013}, we select Eqs. (\ref{eq:fperp_m})-(\ref{eq:pois_bz_m}) and (\ref{eq:pois_ez_m}) to solve for the electromagnetic fields.

Similar to other QSA codes and Darwin model codes \cite{nielson1976}, it is not straightforward to solve the Poisson-like equations and therefore a predictor-corrector iteration is necessary to implicitly determine part of field components. The difficulty in our code arises because the source terms in Eqs. (\ref{eq:pois_br_m}) and (\ref{eq:pois_bphi_m}) are not known at the appropriate time step.  We use the same time indexing as in QuickPIC \cite{huang2006,an2013}. The momentum $\bm{p}$ and Lorentz factor $\gamma$ for the plasma particles are defined on integer time steps, $\xi=n_\xi\Delta_\xi$, while the transverse position $\bm{x}_\perp$ and all the Eulerian quantities including $\psi^m$, $\bm{E}^m$, $\bm{B}^m$, $(\rho-J_z)^m$, $\bm{J}^m$ and $\pp{\bm{J}_\perp^m}{\xi}$ are defined on half-integer time steps, $\xi=(n_\xi+\hf)\Delta_\xi$. In order to deposit $\pp{\bm{J}_\perp^m}{\xi}$ and $\bm{J}^m$, the momentum $\bm{p}^{n_\xi+\hf}$ (the superscript denotes the index of $\xi$) needs to be known. These could be obtained by averaging  $\bm{p}^{n_\xi+1}$ and $\bm{p}^{n_\xi}$ but $\bm{p}^{n_\xi+1}$ is not known because the fields at $\xi=(n_\xi+\hf)\Delta_\xi$  are not known.  Therefore, an iteration procedure is needed.
The $\bm{B}^m$ and $\bm{E}^m$ solved at $\xi=(n_\xi-\hf)\Delta_\xi$ are used as an appropriate initial guess at $\xi=(n_\xi+\hf)\Delta_\xi$.
These are then used to predict $\bm{p}^{n_\xi+1}$ in a leapfrog manner and the $\bm{p}^{n_\xi+\hf}$ is simply evaluated by the average $(\bm{p}^{n_\xi}+\bm{p}^{n_\xi+1})/2$, which we call the predictor procedure. We note that as described in ref. \cite{an2013}, $\pp{\bm{J}_\perp^m}{\xi}$ is obtained by analytically evaluating the derivative of the shape function and not through a finite difference operation of $\bm{J}_\perp^m$. Using this method the particle positions do not need be updated within the predictor procedure. The predicted $\bm{p}^{n_\xi+\hf}$ are then used to deposit the source terms $\pp{\bm{J}_\perp^m}{\xi}$ and $\bm{J}^m$  which are used to  improve  the values of $\bm{B}^m$ and $\bm{E}^m$ from the initial guesses/predictions. This operation is called corrector procedure. To guarantee the procedure for correcting $\bm{B}^m_\perp$ is stable and that it converges, an iterative form of the Poisson equation is used
\begin{align}
\Laplace_m B_r^{m,l+1}-\left(1+\frac{1}{r^2}\right)B_r^{m,l+1}-\frac{2\ii m}{r^2}B_\phi^{m,l+1}&=-\left(\PP{J_\phi^m}{\xi}\right)^l-\frac{\ii m}{r}J_z^{m,l}-B_r^{m,l}, \nonumber \\
\Laplace_m B_\phi^{m,l+1}-\left(1+\frac{1}{r^2}\right)B_\phi^{m,l+1}+\frac{2\ii m}{r^2}B_r^{m,l+1}&=\left(\PP{J_r^m}{\xi}\right)^l+\PP{J_z^{m,l}}{r}-B_\phi^{m,l}, \nonumber
\end{align}
where the superscript $l$ denotes the iteration step. The other components of the fields, $\bm{E}_\perp^m$, $B_z^m$, $E_z^m$ can then be obtained once $\bm{B}_\perp$ is known via Eqs. (\ref{eq:fperp_m}), (\ref{eq:pois_bz_m}) and ($\ref{eq:pois_ez_m}$) respectively (note that $\psi^m$ is already known before the predictor-corrector iteration). This predictor-corrector iteration can be conducted for an arbitrary number of times until the answers are convergence to a desired accuracy.

Unlike in 3D QuickPIC where the equations for the two components of the $\bm{B}_\perp$ are decoupled, Eqs. (\ref{eq:pois_br_m}) and (\ref{eq:pois_bphi_m}) are coupled. For numerical reasons, we instead seek solutions to a set of decoupled equations by introducing new variables $B_+^m=B_r^m+\ii B_\phi^m$ and $B_-^m=B_r^m-\ii B_\phi^m$ in QPAD. With these new field variables, the decoupled equations can be written as
\begin{equation}
\label{eq:pois_decouple_m_iter}
\left(\PP{^2}{r^2}+\frac{1}{r}\PP{}{r}-\frac{(m\pm1)^2}{r^2}-1\right)B_\pm^{m,l+1}=S_\pm^{m,l}-B_\pm^{m,l}
\end{equation}
where
\begin{equation}
S_\pm^m=-\PP{J_\phi^m}{\xi}-\frac{\ii m}{r}J_z^m\pm\ii\left(\PP{J_r^m}{\xi}+\PP{J_z^m}{r}\right). \nonumber
\end{equation}
As we will see in the next section, after discretization, the decoupled equations become tri-diagonal linear systems for which the efficient cyclic reduction algorithm \cite{press2010} can be applied. On the other hand, the original coupled equations would be solved using classic iterative methods or sparse matrix techniques which typically are computationally less efficient.

For computationally simplicity, in the azimuthal mode expansion method, we treat the fields from the beam separately. Due to the approximation that the transverse current $\bm{J}_\perp$ is negligible for beam particles and these particles travel at a speed very closed to the speed of light, $c$ it follows that $\rho^m_\text{beam}\simeq J_{z,\text{beam}}^m$. There is thus no transverse current from the beam which implies that longitudinal fields $B_z^m$ and $E_z^m$ from the beam vanish, and that Eqs. (\ref{eq:pois_br_m}) and (\ref{eq:pois_bphi_m}) reduce to an electrostatics problem,
\begin{equation}
\bm{B}^m_{\perp,\text{beam}}=\bm{e}_r\frac{\ii m}{r}A_z^m+\bm{e}_\phi\PP{A_z^m}{r} \nonumber
\end{equation}
and $A_z^m$ satisfies
\begin{equation}
\label{eq:pois_phi_m}
-\Laplace_m A^m_z=J^m_{z,\text{beam}}=\rho^m_\text{beam}.
\end{equation}
Once  $\bm{B}^m_{\perp,\text{beam}}$ is known then the electric fields can be obtained through $E^m_{r,\text{beam}}=B^m_{\phi,\text{beam}}$ and $E^m_{\phi,\text{beam}}=-B^m_{r,\text{beam}}$.

\section{Algorithm implementation}
\label{sec:algorithm}

\subsection{Numerical workflow in QPAD}
\label{subsec:workflow}

\begin{figure}[htbp]
\centering
\includegraphics[width=\textwidth]{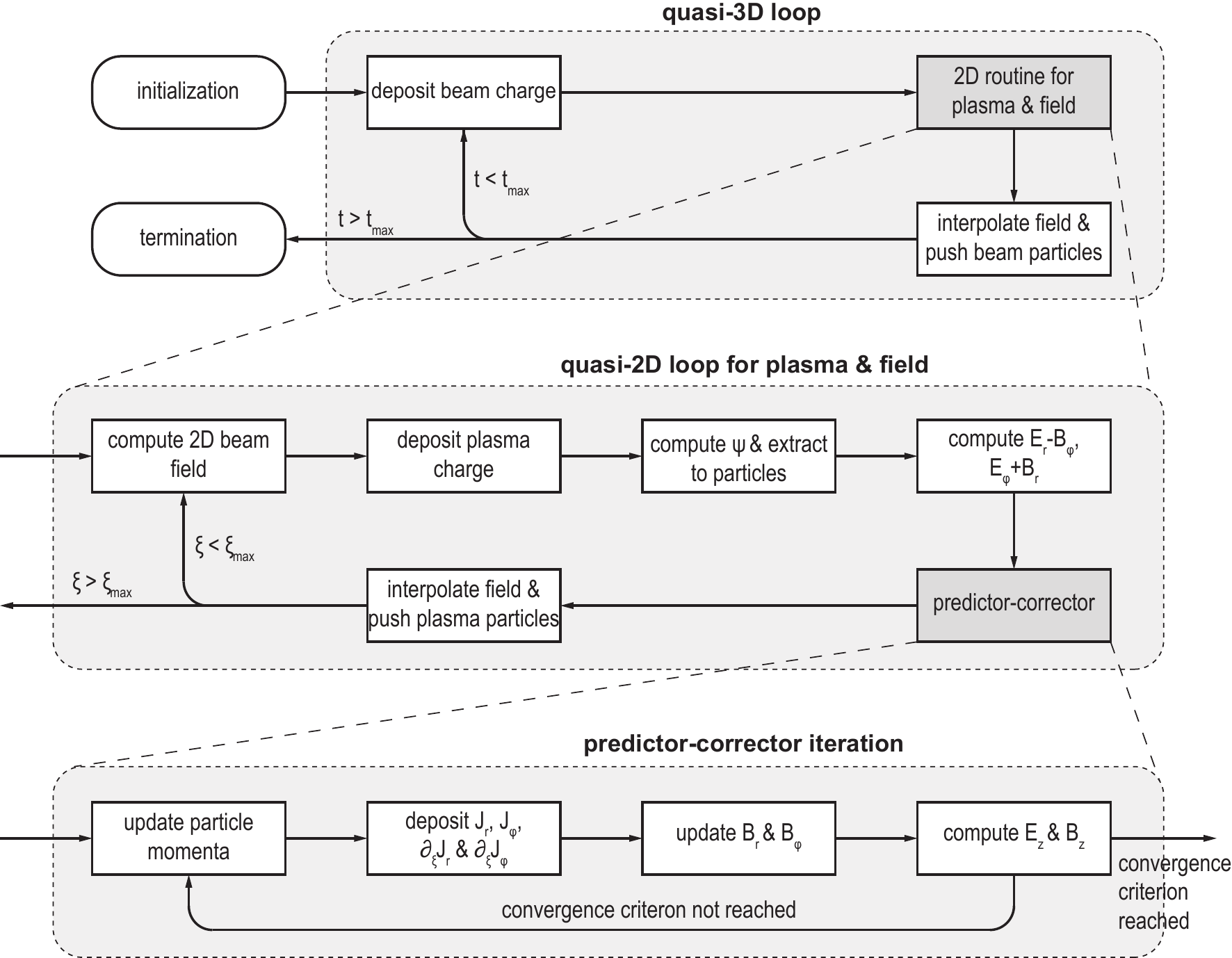}
\caption{The numerical workflow of QPAD}
\label{fig:workflow}
\end{figure}

In this section, we briefly introduce the numerical workflow in QPAD.  QPAD consists of three loops (see figure \ref{fig:workflow}). The outermost level is the the quasi-3D loop in which the charge (current) of the beam particles are deposited onto the $r$-$\xi$ plane for multiple Fourier harmonics, and the beam particles are pushed in $s$ in the full 3D space described by $(x,y,\xi)$ coordinates. The particles are pushed using the  leapfrog method with second-order accuracy.

The quasi-2D loop is embedded into the quasi-3D loop to solve the harmonic amplitudes for all the fields with the plasma and beam charges and currents as sources. The motion of plasma particles is in the 2D space described by $(x, y)$ and particles are pushed in the coordinate $\xi$. In this loop, the evolution of fields and the motion of plasma particles are updated slice by slice along the negative $\xi$-direction. The transverse fields from the particle beam are first calculated at a given slice. This together with the self-consistent fields from the plasma particles are used to advance the particles to new position and momenta at the next slice. In the quasi-static algorithm the particle's charge depends on its speed in the $\hat z$ direction and there are well defined relationships between $\bm p_z$, $\bm p_\perp$ and $\psi$. Therefore, the pseudo-potential $\psi$ must also be interpolated to each particle's position and stored for the subsequent particle push. The equation of motion for a plasma particle is,
\begin{equation}
\DD{\bm{p}_\perp}{\xi}=\frac{q\gamma}{1+\frac{q}{m}\psi}\left[\bm{E}_\perp+\left(\frac{\bm{p}}{\gamma}\times\bm{B}\right)_\perp\right] \nonumber
\end{equation}
and
\begin{equation}
p_z=\frac{1+p_\perp^2-(1-\frac{q}{m}\psi)^2}{2(1-\frac{q}{m}\psi)}. \nonumber
\end{equation}
Once $\psi$ is known, then the transverse fields $E_r-B_\phi$ and $E_\phi+B_r$ can be obtained by taking a transverse gradient of $\psi$ according to Eq. (\ref{eq:fperp_m}). The next step is to call the predictor-corrector iteration to implicitly solve the fields induced by plasma as described earlier. The iteration loop starts with updating the particle momenta by using an initial guess for $\bm{E}$ and $\bm{B}$. The predicted momenta are then used to deposit the source terms $\bm{J}$ and $\pp{\bm{J}_\perp}{\xi}$ needed to solve for $\bm{B}_\perp$. The updated $\bm{E}_\perp$ is evaluated by subtracting $\bm{B}_\perp$ from $-\nabla_\perp\psi$ according to Eq. (\ref{eq:fperp_m}). With the updated $\bm{J}$, the longitudinal field components $E_z$ and $B_z$ can be straightforwardly solved using Eqs. (\ref{eq:pois_ez_m}) and (\ref{eq:pois_bz_m}). This iteration is terminated when a maximum iterative step is reached or the updated fields meet a specified criterion for convergence
\begin{equation}
\frac{\max|\bm{B}^{l+1}-\bm{B}^l|}{\max|\bm{B}^l|}<tol. \nonumber
\end{equation}
where the chosen tolerance is typically $10^{-3}$ or smaller. The last step in the quasi-2D loop is pushing the plasma particles with the converged electromagnetic fields (interpolating the fields onto the particle position) and $\psi$ previously stored on each particle.

In QPAD, the position and momenta of plasma particles are advanced using the classic Boris integrator \cite{boris1972,qin2013} in a leapfrog manner. All the fields in QPAD are solved using  second-order accurate finite difference methods in conjunction with the multi-grid method. A finite difference version of free boundary conditions for different types of field components are implemented into QPAD as well. New current and charge deposition schemes needed for cylindrical geometry and azimuthal decomposition are also developed and implemented. In the next two sections, we describe the numerical implementation of the field solver and deposition in detail.

\subsection{Finite difference Poisson field solver}

In 3D quasi-static PIC codes based on Cartesian coordinates, e.g., QuickPIC and HiPACE \cite{mehrling2014} the fields are solved using FFTs as they are fast and accurate. High parallel scalability is obtained through careful considerations on minimizing data transfer and the use of a pipelining algorithm \cite{feng2009,an2013}. In QPAD, we adopt finite difference (FD) methods to solve Poisson equations because FFTs can no longer be directly used in cylindrical geometry. We define all the fields on the integer grid points $r_i$, \emph{i.e.} $r_i=i\Delta_r$ for $i=1,\dots,N$ where $N$ is the total number of grid points in $\hat{r}$-direction and $\Delta_r$ is the radial cell size. Using a 3-point discretization, the $\Laplace_m$ operator with second-order precision can be written as
\begin{equation}
\Laplace_m U\rightarrow
\beta_i^- U_{i-1}-\alpha_i U_i+\beta_i^+ U_{i+1} \nonumber
\end{equation}
where
\begin{equation}
\beta_i^\pm=\frac{1}{\Delta_r^2}\pm\frac{1}{2r_i\Delta_r},\quad
\alpha_i=\frac{2}{\Delta_r^2}+\frac{m^2}{r_i^2} \nonumber
\end{equation}
and $U$ is an arbitrary scalar field. The operator $\pp{}{r}$ is approximated with the a central difference indexing with second-order precision. Similarly, the operator in Eq. (\ref{eq:pois_decouple_m_iter}) is discretized as
\begin{equation}
\left(\PP{^2}{r^2}+\frac{1}{r}\PP{}{r}-\frac{(m\pm1)^2}{r^2}-1\right)U\rightarrow
\beta_i^- U_{i-1}- \mu_i U_i+\beta_i^+ U_{i+1} \nonumber
\end{equation}
with
\begin{equation}
\mu_i=\frac{2}{\Delta_r^2}+\frac{(m\pm1)^2}{r_i^2}+1. \nonumber
\end{equation}

In QPAD, the governing Eqs. (\ref{eq:fperp_m}), (\ref{eq:pois_psi_m}), (\ref{eq:pois_bz_m}), (\ref{eq:pois_ez_m}), (\ref{eq:pois_decouple_m_iter}) and (\ref{eq:pois_phi_m}) are all discretized. These Poisson equations are converted into tri-diagonal linear systems which can benefit from fast solvers using the cyclic reduction method. These solvers are implemented with the library Hypre \cite{falgout2002} developed and maintained by LLNL.


\subsection{Boundary conditions}

Both conducting and free (open) boundary conditions have been implemented in QPAD. Conducting boundary conditions are more standard and are not described further. The basic idea for free or open boundaries is to assume that the space outside the computational domain is vacuum and that it extends to infinity. Therefore, solutions can be obtained by solving a series of Laplace equations. The boundary values can be determined by utilizing the fact that the fields are continuous at the boundary. See \ref{sec:app:free_bnd} for the details of the derivation.

When implementing the field solvers in cylindrical geometry, issues with respect to singularities on the axis are inevitable issue because of the presence of the $1/r$ term. As we discussed above, all the field components in QPAD are defined on integer grid points. Therefore,  all of the Poisson equations can have a singularity at $r=0$. These singularities can be properly treated by considering the properties of different field components at $r=0$. As pointed out by Constantinescu and Lele \cite{constantinescu2002}, for any scalar of component of a field in a Cartesian coordinate directions, \emph{i.e.}, $(\psi, \varphi, E_z, B_z, \rho-J_z, J_z)$, the $m=0$ mode is non-zero at $r=0$ while other modes are zero at $r=0$.
On the other hand, for the field components in cylindrical coordinate directions $(E_r, E_\phi, B_r, B_\phi, J_r, J_\phi)$ the $m=1$ mode is non-zero at $r=0$ and the other modes are zero at $r=0$.

The field components whose boundary values at $r=0$ need to be determined therefore only include $\psi^0$, $\varphi^0$, $B_z^0$, $E_z^0$ and $B_\pm^1$; all other field components vanish at $r=0$. The singularity of the  $\frac{1}{r}\PP{U}{r}$ term on the LHS of each Poisson equation (where $U$ denotes any of fields mentioned above) can be eliminated by applying L'Hospital's rule, so that $\frac{1}{r}\PP{U}{r}\rightarrow\PP{^2U}{r^2}$. The terms having $1/r$ on the RHS of Poisson equation can be treated in the same manner. There is another important property for components in the cylindrical coordinate directions \cite{constantinescu2002}. The combinations $U_r+imU_\phi$ and $U_\phi-imU_r$ vanish at $r=0$ for arbitrary $m$, which implies $B^1_+$ (recalling the definition is $B_+^m\equiv B_r^m+iB_\phi^m$) vanishes at $r=0$ even though both $B_r^1$ and $B_\phi^1$ are non-zero on the axis. For $m\neq1$ modes $B_r^m$ and $B_\phi^m$ are naturally zero at $r=0$ according to previous discussion, therefore, $B_+^m$ vanishes on the axis for arbitrary $m$. Considering the symmetry of different fields around the axis, the discrete version of boundary conditions at $r=0$ can therefore be summarized as follows,
\begin{align}
4(\psi_1^{m=0}-\psi_0^{m=0}) &= -(\rho-J_z)_0^{m=0} \Delta_r^2 \\
2(E_{z,1}^{m=0}-E_{z,0}^{m=0}) &= J_{r,1}^{m=0} \Delta_r \\
2(B_{z,1}^{m=0}-B_{z,0}^{m=0}) &= -J_{\phi,1}^{m=0} \Delta_r \\
4(B_{-,1}^{m=1}-B_{-,0}^{m=1}) &= -\left[\left(\PP{J_\phi}{\xi}\right)_0^{m=1} + i\left(\PP{J_r}{\xi}\right)_0^{m=1} + 2i\frac{J_{z,1}^{m=1}}{\Delta_r} \right] \Delta_r^2
\end{align}

\subsection{Deposition of source terms}

In order to solve the 1D Poisson equations for each harmonic amplitude, the source terms on the RHS of the governing Eqs. (\ref{eq:fperp_m})-(\ref{eq:pois_bz_m}) and (\ref{eq:pois_ez_m}) must be deposited from the particle information (charge, position and momentum) onto the grid points. The source terms to be deposited include $\rho^m-J_z^m$, $\bm{J}^m$ and $\pp{\bm{J}_{\perp}^m}{\xi}$. Since these source terms are defined on the grid in the $\hat{r}$-direction while the particles are described by Cartesian coordinates, we need to transform the particle positions and momenta from the cylindrical to Cartesian coordinates in QPAD. The following equation is used to deposit the current as in QuickPIC,
\begin{equation}
\label{eq:dep_jay}
\bm{J}=\frac{1}{\text{Vol.}}\sum_i\frac{q_i\bm{v}_i}{1-v_{iz}}S(\bm{x}_\perp-\bm{x}_{i\perp})=\frac{1}{\text{Vol.}}\sum_i\frac{q_i\bm{p}_i}{1-\frac{q_i}{m_i}\psi_i}S(\bm{x}_\perp-\bm{x}_{i\perp})
\end{equation}
where $S(\bm{x}_\perp-\bm{x}_{i\perp})$ is the particle shape function to interpolate the particle quantities at $i$th particle's transverse position $\bm{x}_{i\perp}$ onto the grid position $\bm{x}_\perp$. The pseudo-potential felt by an individual particle $\psi_i$ is obtained by interpolating the $\psi$ solved on the grid to the position of the particle. The second expression for $\bm J$ can be obtained by multiplying the numerator and denominator of the first expression by the Lorentz factor $\gamma_i$, and using the constant of motion under the QSA, $\gamma-p_z=1-(q/m)\psi$. In order to derive the deposition scheme in QPAD in which the azimuthal direction is gridless, we  expand $S(\bm{x}_\perp-\bm{x}_{i\perp})$ into a Fourier series as well. In cylindrical geometry, the interpolation function is defined as
\begin{equation}
S(\bm{x}_\perp-\bm{x}_{i\perp})\equiv
\frac{1}{r}S_r(r-r_i)S_\phi(\phi-\phi_i), \nonumber
\end{equation}
which is subject to the normalization condition $\int dr d\phi\ S_rS_\phi=1$. Next, we expand $S_\phi$ into azimuthal harmonics
\begin{equation}
S_\phi(\phi-\phi_i)=\sum_m S_\phi^m(\phi_i)e^{\ii m\phi} \nonumber
\end{equation}
where
\begin{equation}
S_\phi^m(\phi_i)=\frac{1}{2\pi}\int_0^{2\pi}d\phi'\ S_\phi(\phi'-\phi_i)e^{-\ii m\phi'}, \nonumber
\end{equation}
and require both $S_r$ and $S_\phi$ to satisfy the normalization condition $\int drS_r=1$ and $\int d\phi S_\phi=1$. If we take $S_\phi$ to be a Dirac delta function(which we do in QPAD), then $S_\phi^m=\frac{1}{2\pi}e^{-\ii m\phi_i}$. In addition,  the current $\bm{J}$ defined on the {\it r-z} grid can be expanded as
\begin{equation}
\bm{J}(r,\phi)=\sum_m \bm{J}^m(r) e^{\ii m\phi}, \nonumber
\end{equation}
the deposition for $\bm{J}^m$ is found to be
\begin{equation}
\bm{J}^m=\frac{1}{\text{Vol.}}\sum_i\frac{q_i\bm{p}_i}{1-\frac{q_i}{m_i}\psi_i}\frac{1}{r}S_r(r-r_i)S^m_\phi(\phi_i). \nonumber
\end{equation}
Therefore, it is actually not necessary to calculate each {\it m} mode but only the $m=0$ mode from each particle. Any $m>0$ mode for an individual particle can be obtained from the $m=0$ contribution by simply multiplying by a phase factor through the relation $\bm{J}^m=\bm{J}^0e^{-\ii m\phi_i}$ or recursively through $\bm{J}^m=\bm{J}^{m-1}e^{-\ii\phi_i}$ if $S(\phi-\phi_i)=\delta(\phi-\phi_i)$ is used.

Likewise, according to ref. \cite{an2013}, the deposition for $(\rho-J_z)^m$ can be written as
\begin{equation}
(\rho-J_z)^m=\frac{1}{\text{Vol.}}\sum_i\frac{q_i}{r}S_r(r-r_i)S_\phi^m(\phi_i), \nonumber
\end{equation}
where $(\rho-J_z)^m=(\rho-J_z)^{m-1}e^{-\ii\phi_i}$ for each particle.

In section \ref{subsec:workflow}, we showed that in the predictor-corrector iteration the source term $\pp{\bm{J}^m_\perp}{\xi}$ at the half-integer time step $\xi=(n+1/2)\Delta_\xi$ needs to be calculated. This can be done in two ways. The first method, which was adopted in the original version of QuickPIC \cite{huang2006}, is to predict $J^m_r$ and $J_\phi^m$ at the next integer time step $\xi=(n_\xi+1)\Delta_\xi$ and approximate the derivative using the centered difference $\pp{J_{r,\phi}^m}{\xi}\vert^{n_\xi+\hf}=(J_{r,\phi}^m\vert^{n_\xi+1}-J_{r,\phi}^m\vert^{n_\xi})/\Delta_\xi$. However, this approach requires repartitioning the particles within a single pass through the iteration loop when using domain decomposition as it requires updating the particle positions and storing previous and predicted values. In the current version of QuickPIC \cite{an2013}, this approach is replaced by analytically calculating the derivative of the current in terms of $\bm{x}_\perp$, $\bm{p}_\perp$ and $\psi$ using their particle shapes, which allows direct deposition without the computationally expensive particle repartitioning procedure. In QPAD, we use the approach in the current version of  QuickPIC to deposit $\pp{\bm{J}_\perp^m}{\xi}$. By definition, we have
\begin{equation}
\begin{aligned}
\label{eq:dep_djrdxi}
\PP{J_r^m}{\xi}&=\frac{1}{\text{Vol.}}\sum_i \PP{}{\xi}\left(\frac{q_ip_{r,i}}{1-\frac{q_i}{m_i}\psi_i}\frac{1}{r}S_rS_\phi^m\right) \\
&=\frac{1}{\text{Vol.}}\sum_i \frac{q_i}{r}\left(\frac{\dd{p_{r,i}}{\xi}}{1-\frac{q_i}{m_i}\psi_i}S_rS_\phi^m+\frac{p_{r,i}\dd{(\frac{q}{m_i}\psi_i)}{\xi}}{(1-\frac{q_i}{m_i}\psi_i)^2}S_rS_\phi^m+\frac{p_{r,i}}{1-\frac{q_i}{m_i}\psi_i}\PP{(S_rS_\phi^m)}{\xi}\right).
\end{aligned}
\end{equation}
It should be pointed out that the $\psi_i$ in the denominator is the total value which is obtained by summing all the harmonics. The derivative of $\psi_i$ with respect to $\xi$ is calculated by
\begin{equation}
\label{eq:dpsidxi}
\DD{\psi_i}{\xi}=E_{z,i}+\PP{\psi_i}{r_i}\DD{r_i}{\xi}+\PP{\psi_i}{\phi_i}\DD{\phi_i}{\xi}
\end{equation}
where the terms $E_{z,i}$, $\PP{\psi_i}{r_i}$ and $\PP{\psi_i}{\phi_i}$ are regarded as the interpolated value of $E_z$, $\PP{\psi}{r}$ and $\PP{\psi}{\phi}$ at the particle's position $(r_i,\phi_i)$. The terms $\DD{r_i}{\xi}$ and $\DD{\phi_i}{\xi}$ are evaluated by
\begin{equation}
\DD{r_i}{\xi}=\frac{p_{r,i}}{1-\frac{q_i}{m_i}\psi_i},\quad
\DD{\phi_i}{\xi}=\frac{1}{r_i}\frac{p_{\phi,i}}{1-\frac{q_i}{m_i}\psi_i}. \nonumber
\end{equation}
For the last term in the bracket of Eq. (\ref{eq:dep_djrdxi}), $\pp{(S_rS_\phi^m)}{\xi}$ is calculated by
\begin{equation}
\begin{aligned}
\PP{}{\xi}\left(S_r(r-r_i)S_\phi^m(\phi_i)\right)&=-\DD{r_i}{\xi}\PP{S_r}{r}S_\phi^m+\DD{\phi_i}{\xi}\PP{S_\phi^m}{\phi_i}S_r \\
&=-\frac{p_{r,i}}{1-\frac{q_i}{m_i}\psi_i}\PP{S_r}{r}S_\phi^m+\frac{p_{\phi,i}}{1-\frac{q_i}{m_i}\psi_i}\frac{1}{r_i}\PP{S_\phi^m}{\phi_i}S_r \\
&=-\frac{e^{-\ii m\phi_i}}{2\pi}\left(\frac{p_{r,i}}{1-\frac{q_i}{m_i}\psi_i}\PP{S_r}{r}+\frac{p_{\phi,i}}{1-\frac{q_i}{m_i}\psi_i}\frac{\ii mS_r}{r_i}\right). \nonumber
\end{aligned}
\end{equation}
where we have applied $S_\phi=\delta(\phi-\phi_i)$ again. Substituting these expressions into Eq. (\ref{eq:dep_djrdxi}), we finally obtain the deposition for $\pp{J_r^m}{\xi}$
\begin{equation}
\begin{aligned}
\PP{J_r^m}{\xi}&=\frac{1}{2\pi\text{Vol.}}\left\{\sum_i q_ie^{-\ii m\phi_i}\left(\frac{\dd{p_{r,i}}{\xi}}{1-\frac{q_i}{m_i}\psi_i}+\frac{p_{r,i}\dd{(\frac{q}{m}\psi_i)}{\xi}}{(1-\frac{q_i}{m_i}\psi_i)^2}-\frac{p_{r,i}p_{\phi,i}}{(1-\frac{q_i}{m_i}\psi_i)^2}\frac{\ii m}{r_i}\right.\right.\\
&\left.\left.-\frac{p_{r,i}^2}{(1-\frac{q_i}{m_i}\psi_i)^2}\frac{1}{r}\right)\frac{S_r}{r}-\PP{}{r}\left(\sum_i q_i e^{-\ii m\phi_i}\frac{p_{r,i}^2}{(1-\frac{q_i}{m_i}\psi_i)^2}\frac{S_r}{r}\right)\right\}
\end{aligned}, \nonumber
\end{equation}
and likewise we can derive the deposition formula for $\pp{J_\phi^m}{\xi}$
\begin{equation}
\begin{aligned}
\PP{J_\phi^m}{\xi}&=\frac{1}{2\pi\text{Vol.}}\left\{\sum_i q_ie^{-\ii m\phi_i}\left(\frac{\dd{p_{\phi,i}}{\xi}}{1-\frac{q_i}{m_i}\psi_i}+\frac{p_{\phi,i}\dd{(\frac{q}{m}\psi_i)}{\xi}}{(1-\frac{q_i}{m_i}\psi_i)^2}-\frac{p_{\phi,i}^2}{(1-\frac{q_i}{m_i}\psi_i)^2}\frac{\ii m}{r_i}\right.\right.\\
&\left.\left.-\frac{p_{r,i}p_{\phi,i}}{(1-\frac{q_i}{m_i}\psi_i)^2}\frac{1}{r}\right)\frac{S_r}{r}-\PP{}{r}\left(\sum_i q_i e^{-\ii m\phi_i}\frac{p_{r,i}p_{\phi,i}}{(1-\frac{q_i}{m_i}\psi_i)^2}\frac{S_r}{r}\right)\right\}.
\end{aligned}
\end{equation}

\section{Simulation results}
\label{sec:simulation}

In this section, we present a small sample of benchmark tests for QPAD compared against results from QuickPIC and 3D OSIRIS. These benchmarks are related to the plasma wakefield accelerator (PWFA) concept which uses high-energy particle beams to excite a plasma wave wake. The plasma wake provides very large accelerating and focusing forces as compared with conventional accelerator structures. These fields can be used to accelerate and/or focus a trailing beam riding on an appropriate phase inside the wake. We present benchmarks for driving wakefields in both the linear and nonlinear regimes with only a single mode (only $m=0$ mode). We also present a benchmark for a case where a second witness beam is placed inside a nonlinear wakefield \cite{lu2006,rosenzweig1991} with an offset in one direction with respect to the drive beam. This leads to a hosing instability \cite{whittum1991,huang2007} and requires keeping at least the $m=1$ mode.

\subsection{Plasma wakefield excitation}

We start by simulating linear wakefield excitation. The linear regime refers to the case that the peak density of the drive beam $n_b$ is much smaller than the background plasma density $n_p$, so that the drive beam only introduces a weak perturbation to the plasma and the background electrons oscillate in a nearly sinusoidal fashion. In this case, the drive beam has a bi-Gaussian density profile with a spot size $k_p\sigma_r=2.0$, bunch length $k_p\sigma_z=0.5$, and peak density $n_b/n_p=0.1$, where $k_p^{-1}$ is the plasma skin depth where $n_b=\frac{N}{(2\pi)^{3/2}} \exp[-(\frac{r^2}{2\sigma^2_r}+\frac{z^2}{2\sigma^2_z})]$ and $N$ is the number of particles in the bunch. Since this scenario possesses azimuthal symmetry, we only include the $m=0$ mode in QPAD which is equivalent to a 2D r-z simulation using codes such as WAKE or LCODE. In the QuickPIC and OSIRIS simulations, the cell size is $\Delta_x=\Delta_y=0.0234\ k_p^{-1}, \Delta_z=0.0195\ k_p^{-1}$. In the QPAD simulation, $\Delta_r=0.0234\ k_p^{-1}, \Delta_z=0.0195\ k_p^{-1}$. The drive beams are initialized with $128\times128\times256$ particles in $x, y$ and $z$ for the QuickPIC simulation and with $128\times32\times256$ particles in $r,\phi$ and $z$ for the QPAD simulation. For the plasma, we use $2\times2$ particles per 2D cell in QuickPIC and uniformly distribute $2\times32$ particles within a ring of width $\Delta_r$ in QPAD. In OSIRIS, $2\times2\times2$ particles per 3D cell are used to initialize both the plasma and beam.

\begin{figure}[htbp]
\centering
\includegraphics[width=\textwidth]{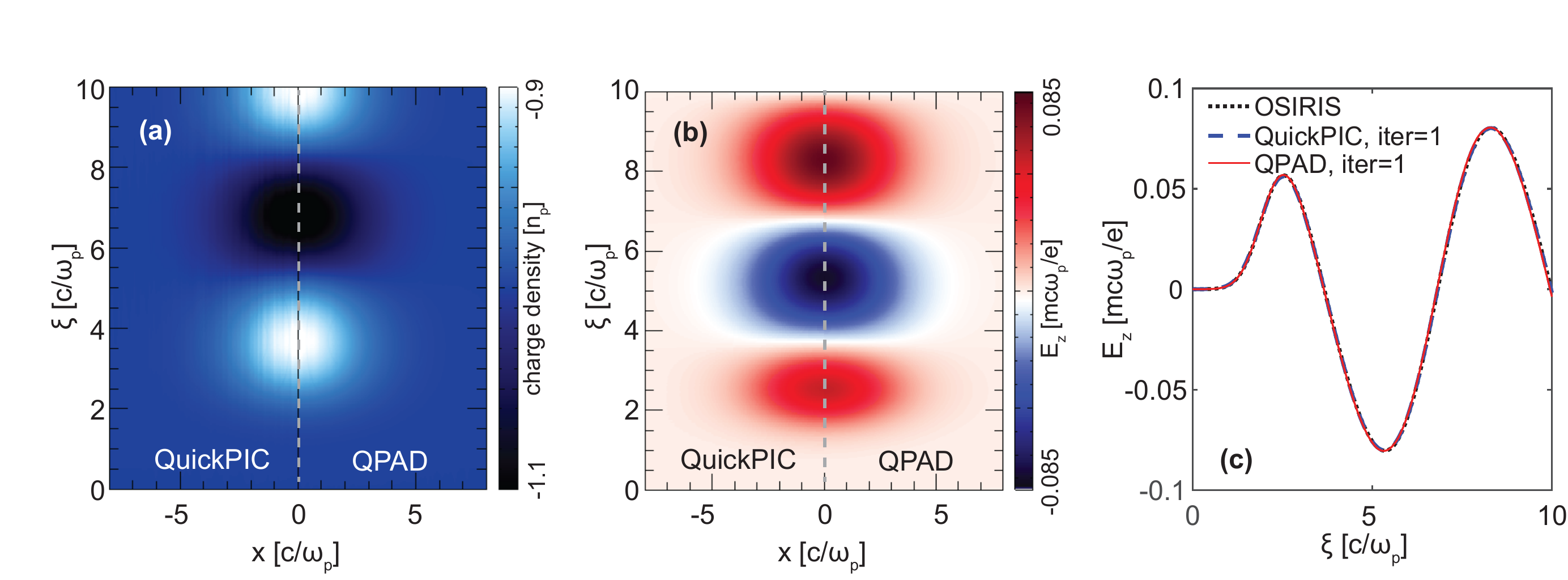}
\caption{Comparison of beam-driven wakefield in linear regime between OSIRIS, QuickPIC and QPAD. (a) Background electron density. (b) $E_z$ field. (c) On-axis lineouts of $E_z$ fields from OSIRIS, QuickPIC and QPAD.}
\label{fig:linear_wake}
\end{figure}

The simulation results are shown in Fig. \ref{fig:linear_wake}. In figure \ref{fig:linear_wake}(a) and (b), we compare the plasma electron density and $E_z$ field between QuickPIC and QPAD runs. The drive beams, whose centers reside at $\xi=2$, move downward and are not displayed in these figures. Figure \ref{fig:linear_wake}(c) compares the lineouts of $E_z$ on the $r=0$  axis between QPAD, QuickPIC and OSIRIS. Here, only one predictor-corrector iteration is conducted in QPAD and this already gives excellent agreement with QuickPIC and OSIRIS. We also conducted convergence tests for the predictor-corrector loop by iterating 1, 3 and 5 times. We found in this scenario, the predictor-corrector loop converges so rapidly that only one iteration is sufficient to reach the desired simulation accuracy.

\begin{figure}[htbp]
\centering
\includegraphics[width=\textwidth]{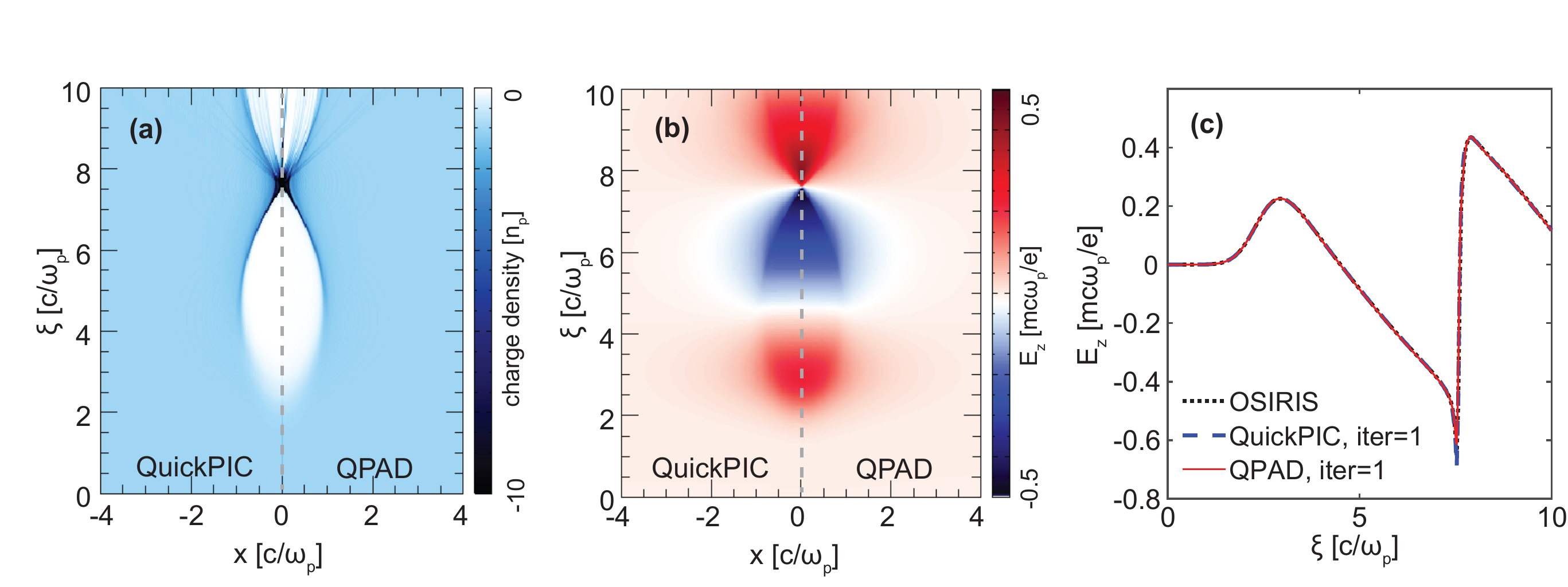}
\caption{Comparison of beam-driven wakefield in nonlinear regime between OSIRIS, QuickPIC and QPAD. (a) Background electron density. (b) $E_z$ field. (c) On-axis lineouts of $E_z$ fields from OSIRIS, QuickPIC and QPAD.}
\label{fig:nonlinear_wake}
\end{figure}

Next, we simulate drive beam parameters for which a nonlinear plasma wakefield is excited. In this case the peak density of the beam is much larger than the plasma density, \emph{i.e.}, $n_b\gg n_p$.
Here, we show an example for which $n_b/n_p=4$, $k_p\sigma_r=0.25$, $\Lambda\equiv(n_b/n_p)(k_p\sigma_r)^2=0.25$ and keep other numerical parameters the same as those in the linear regime case. In the nonlinear regime, the $E_z$ on axis now looks similar to a sawtooth wave as shown in fig. \ref{fig:nonlinear_wake}(c). In the region where the background plasma electrons are fully evacuated by the drive beam (from $\xi=3$ to $\xi=7$), the $E_z$ field almost drops linearly to its minimum at the rear of the first ion bubble. From fig. \ref{fig:nonlinear_wake}, we can see that QPAD with only one predictor-corrector iteration still gives results in almost perfect agreement with OSIRIS and QuickPIC. Similarly to the convergence test for the linear regime, the predictor-corrector iteration is found to converge rapidly. Running the iteration more than once does not make an observable difference to the simulation results.

\begin{figure}[htbp]
\centering
\includegraphics[width=\textwidth]{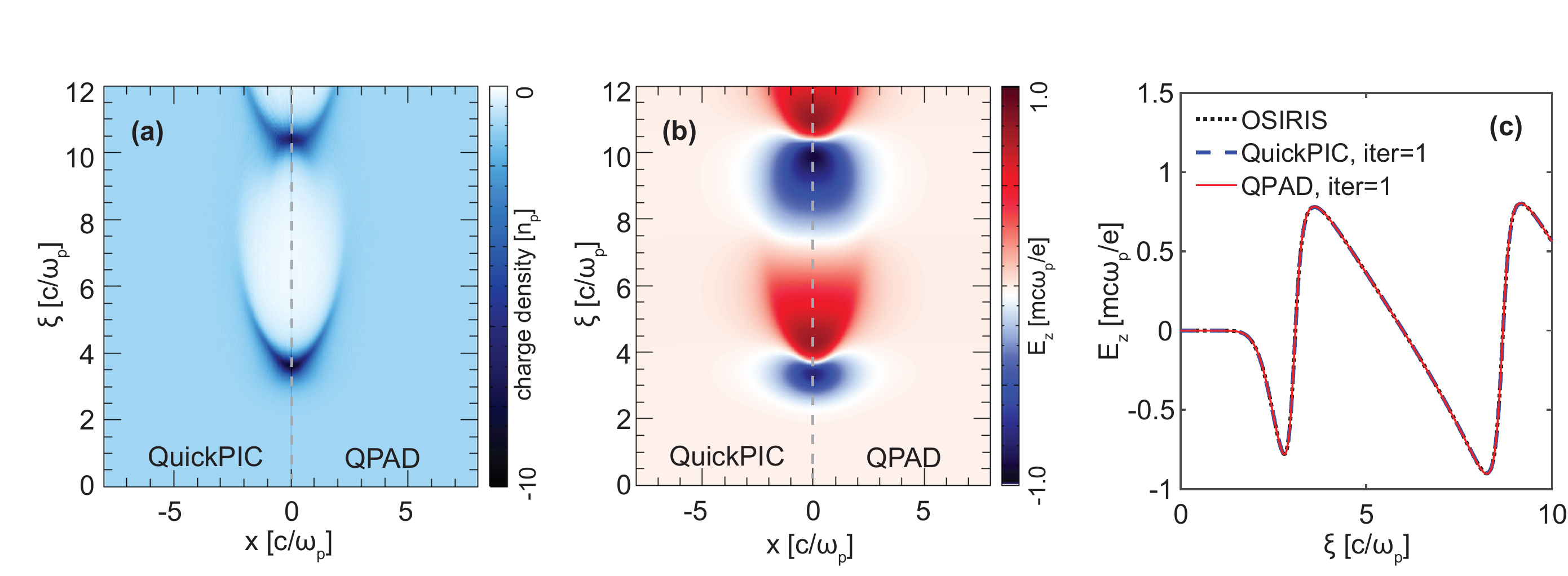}
\caption{Comparison of positron-beam-driven wakefield between OSIRIS, QuickPIC and QPAD. (a) Background electron density. (b) $E_z$ field. (c) On-axis lineouts of $E_z$ fields from OSIRIS, QuickPIC and QPAD.}
\label{fig:positron}
\end{figure}

Besides an electron beam, a very short positron or proton beam can also excite a bubble-like plasma wake. Due to the attractive force from the positron bunch, the background electrons are ``sucked in'' first by the drive beam rather than ``blown out'' as is the case for an electron beam driver. This leads to the background electrons forming a density peak at the front of the first bucket, and the $E_z$ field being negative in that region. After the plasma electrons collapse to the axis, they then overshoot and eventually form a blowout type wake in the second wavelength. In figure \ref{fig:positron}(a) and (b), a bi-Gaussian positron beam with $n_b/n_p=2.5,\ k_p\sigma_r=0.8,\ k_p\sigma_\xi=0.46$ and the center resides at $\xi=3$ moves downward. Again, we use only one predictor-corrector iteration to achieve good agreements with the results of QuickPIC and OSIRIS.

\subsection{Hosing instability}

\begin{figure}[htbp]
\centering
\includegraphics[width=0.8\textwidth]{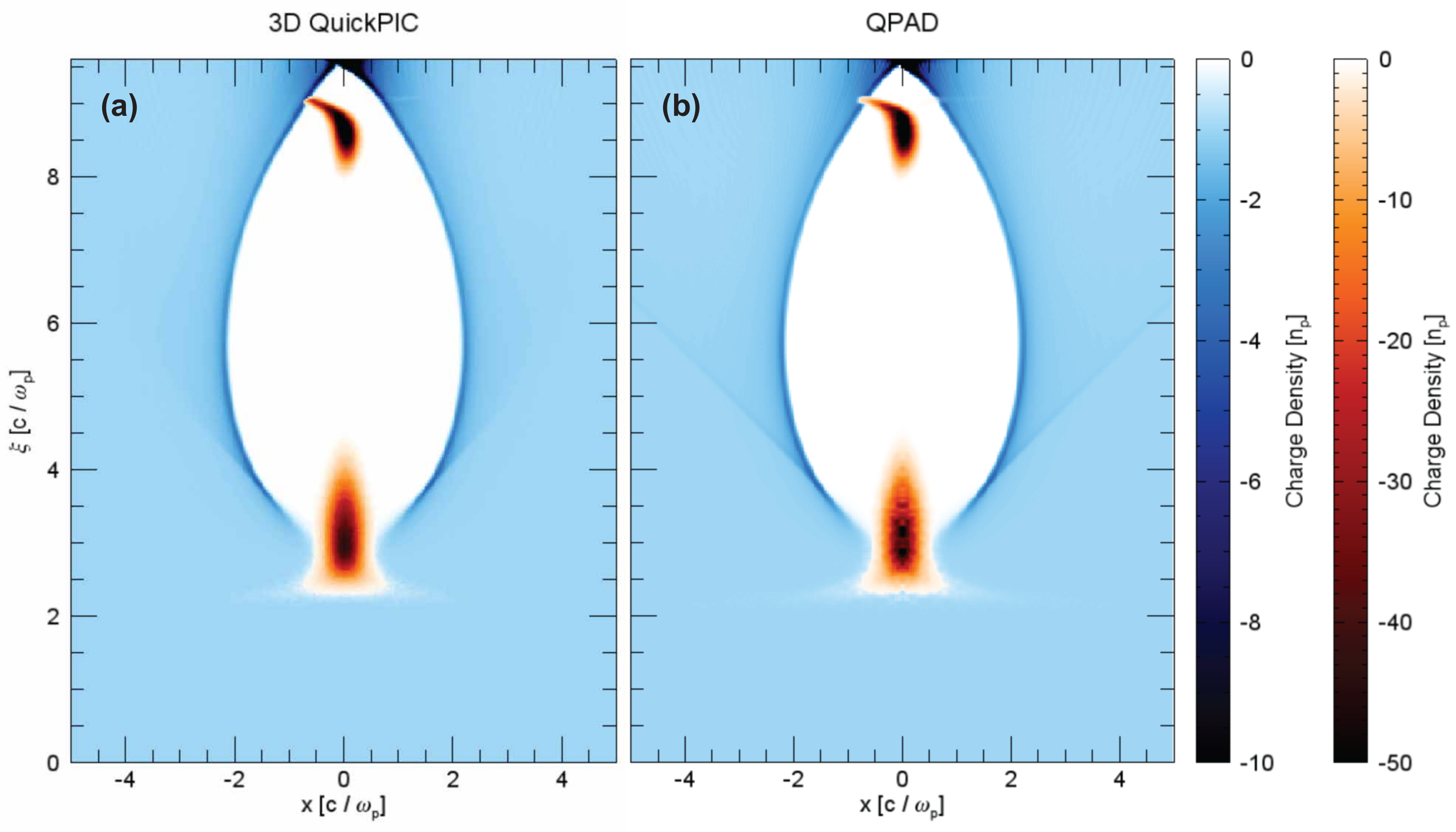}
\caption{Density distribution of plasma electrons and beams in (a) full 3D QuickPIC and (b) QPAD simulations.}
\label{fig:hosing_density}
\end{figure}

In this section, we present a simulation of what is called the hosing instability in PWFA \cite{huang2007}. The hosing instability is one of the major impediments for PWFA and can lead to beam breakup. Although an azimuthally symmetric r-z code such as WAKE and LCODE is very efficient to model PWFA, it cannot be used to investigate the physics involving asymmetries such as the hosing instability. For hosing we only compare QPAD against QuickPIC. The drive beam has a bi-Gaussian profile with a peak density $n_b/n_p=93.5$, an rms spot-size $k_p\sigma_r=0.14$ and an rms bunch length $k_p\sigma_z=0.48$ which corresponds to $\Lambda\simeq1.8$. The trailing beam parameters are $n_b/n_p=56$, $k_p\sigma_r=0.14$ and $k_p\sigma_z=0.24$.  For both the plasma and the beams there are 16 macro-particles distributed in $\phi$ while for the plasma there are 2 macro-particles per r-z cell. Within the region $[-5\sigma_r,+5\sigma_r]\times[-5\sigma_z,+5\sigma_z]$ the drive beam and trailing beam have $128\times512$ and $128\times256$ particles respectively, and have 16 particles azimuthally. The drive beam is initialized axisymmetrically while the trailing beam has a small centroid offset of $0.038\ k_p^{-1}$ in $x$-direction. For the full 3D QuickPIC simulation, the plasma has $2\times2\times2$ particles per cell and the drive beam and trailing beam have $128\times128\times512$ and $128\times128\times256$ particles within the $5\sigma$ rectangular block. The initial longitudinal proper velocity corresponds to $\gamma \beta_z=20000$ for both the drive and trailing beams.  In the QPAD simulation, modes $m=0, 1, 2$, and 3 are included.  Figure \ref{fig:hosing_density} shows the density distribution with the background plasma electrons and beams colored blue and red respectively. The snapshots were taken at $\omega_p t=20000$. It can be seen that there is excellent agreement between QPAD and QuickPIC for the motion of the trailing beam even for this nonlinear problem.

\begin{figure}[htbp]
\centering
\includegraphics[width=0.8\textwidth]{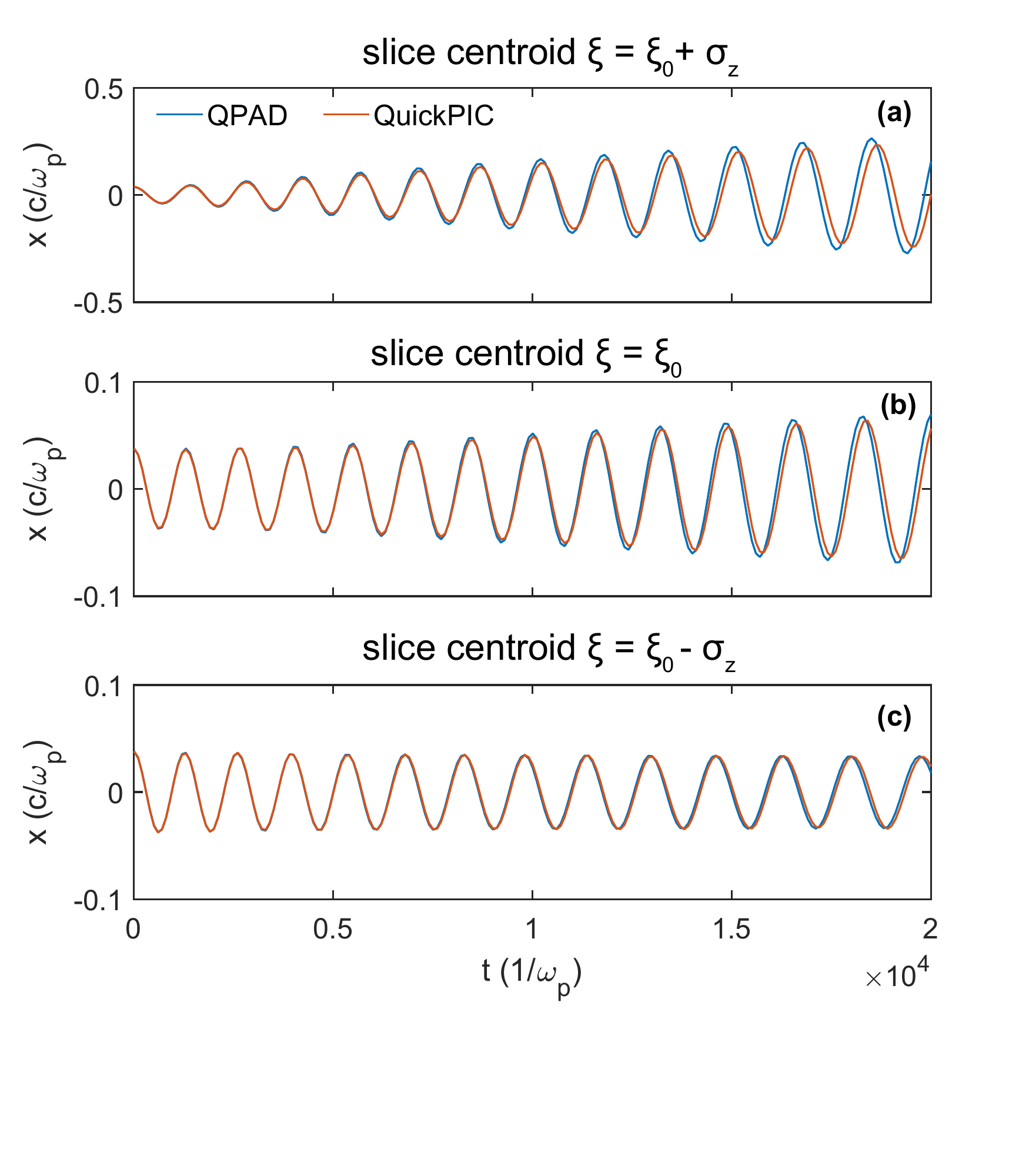}
\caption{Beam centroid oscillation of slice residing at (a) $+\sigma_z$, (b) 0 and (c) $-\sigma_z$ with respect to the beam center $\xi_0$.}
\label{fig:hosing_centroid}
\end{figure}

A more careful comparison between the hosing results is obtained by investigating the beam centroid oscillation during the entire acceleration distance for different beam slices. Figure \ref{fig:hosing_centroid}(a)-(c) plots the centroid oscillation for three slices, residing at $+\sigma_z,\ 0$ and $-\sigma_z$ with respect to the beam center $\xi_0$. The centroid is defined as $\frac{1}{N}\sum x_i$ where the sum is taken over all particles within a slice at $z \pm 0.1k_p^{-1}$ and $N$ is the number of particles. The amplitude of the centroid oscillation for the slice closer to the beam head [figure \ref{fig:hosing_centroid}(c)] remains nearly constant in $s$, the amplitude grows in $s$ with a larger growth rate the farther the slice is behind the center of the beam [figure \ref{fig:hosing_centroid}(a) and (b)]. This qualitatively agrees well with the theoretical prediction on the instability growth. Except for a slight phase difference that is evident for larger values of $s$, there is excellent agreement between QPAD and full 3D QuickPIC simulations. These differences may be due to the truncation of the azimuthal mode expansion at $m=3$. We emphasize that a code such as QPAD is also a powerful too for carrying out large parameter scans even if the results are not quantitatively correct.

\section{Algorithm complexity}
\label{sec:performance}

The azimuthal-decomposition-based algorithm has the potential to greatly reduce the computational requirements without much loss in accuracy when modeling 3D physics when the problem only has low order azimuthal asymmetry. This is because it requires fewer grid points and hence few particles. We can make a straightforward estimation of the speedup over a full 3D quasi-static code.

In QuickPIC, the fields are solved on a 2D slab (usually a square) with $n_\text{mesh}=N^2$ grid points, so the cost of the Poisson solver is $O(N^2\log(N))$ assuming the fast FFT method is used. In QPAD, we solve fields on a 1D mesh with $n_\text{mesh}=N/2$ grid points for $2m_\text{max}+1$ components ($m=0$ mode and real/imaginary parts for $m>0$ modes) where $m_\text{max}$ is the index of the highest azimuthal mode that is kept.  Therefore, the cost of the Poisson solver is $(2m_\text{max}+1)O[(N/2)\log(N/2)]$ using the cyclic reduction method. The speedup for the field solve will therefore scale as $\sim O(N)/(m_\text{max}+\hf)$ compared with the FFT method used in QuickPIC. In QuickPIC, a total number of $N^2N_{\text{ppc},x}N_{\text{ppc},y}$ macro-particles for plasma species are used where $N_{\text{ppc},\eta},\ (\eta=x,y)$ denotes the particle number per cell in the $\eta$-direction. In QPAD, there are only $NN_{\text{ppc},r}N_{p,\phi}/2$ macro-particles for each plasma species, where $N_{\text{ppc},r}$ is number of particles per $r$-$z$ cell  and $N_{p,\phi}$ is the number of particles distributed over $0<\phi<2\pi$. Assuming the computational cost of pushing particles is proportional to the total macro-particle number, the speedup therefore scales as $2NN_{\text{ppc},x}N_{\text{ppc},y}/(N_{\text{ppc},r}N_{p,\phi})\sim O(N)$. For a majority of PWFA problems, the configuration with $m_\text{max}\leq2$ and particle number $N_{p,\phi}\sim10,\ N_{\text{ppc},r}\sim N_{\text{ppc},x}$ or $N_{\text{ppc},y}$ are enough to capture the dominant azimuthal asymmetry to effectively simulate the physics with nearly round drive beams, so that considerable speedup can be achieved for typical numerical parameters. The goal of this paper is to describe how to implement an azimuthal mode expansion into a quasi-static PIC code. Issues with respect to optimization will be addressed in future publications. The parallelization in QPAD is also similar to that in QuickPIC. The code is parallelized using MPI to run on distributed memory clusters, which is implemented by means of spatial decomposition in r and z dimensions. However, owing to the basic numerical scheme of a quasi-static code, the parallelization in $r$ direction differs essentially from the that in $\xi$ direction. The parallelization in $r$ is similar to that in full explicit PIC codes with the macro-particles exchange between neighboring processors. The interprocess exchange of field values at the domain boundaries is handled by the built-in routines of Hypre library. In the $\xi$ direction, we use pipelining algorithm to allow the transverse process slabs to run asynchronously, which can significantly inhibit the idle time.

\section{Conclusion}

We have describe QPAD, a new quasi-static PIC code that uses the azimuthal Fourier decomposition for the fields. The new code utilizes the workflow and routines of QuickPIC in which a 2D code for evolving the plasma particles in a time like variable $\xi$ is embedded into a 3D code that advances beam particles in a time like variable $s$. In QPAD, all the field components are decomposed into a few Fourier harmonics in $\phi$. In the 2D part of the code each amplitude depends on $r$ and evolves in $\xi$. Therefore, in this part of the code the fields are only defined on a 1D grid in $r$. The quasi-static version of  Maxwell's equations for each harmonic amplitude are therefore one-dimensional, making the new code much faster. A full set of Poisson-like equations that exactly correspond to those used in the full 3D QuickPIC are written in cylindrical geometry. A full set of 1D Poisson equations in $r$ are solved for the Fourier amplitudes in $\phi$ for the relevant fields. To simplify the calculation, we introduced linear combinations of the complex amplitudes, $B^m_\pm$, to decouple the equations for $B^m_r$ and $B^m_\phi$. Open (free) boundary conditions are implemented for all the fields. For the particle module, the macro-particles are distributed and advanced in $\xi$ in a 2D space $(r,\phi)$.  A predictor-corrector routine is described. A novel deposition method for $\PP{\bm{J}}{\xi}$, $\bm{J}$ and $\rho-J_z$ for each harmonics is described and implemented. This scheme does not require updating the particle positions to obtain $\PP{\bm{J}}{\xi}$ which reduces the complexity of the predictor corrector routine.  The new code was benchmarked and compared against results from 3D OSIRIS and QuickPIC for a few sample cases. Excellent agreement was found for both wake excitation of plasma wave wakes from particle beam drivers (electrons and positrons) and for the electron hosing instability. Directions for future work include optimizing the field solver to reduce the across-node data communication, adding multi-threading features (OpenMP), and implementing more physics including field-ionization, radiation reaction, and the ponderomotive guiding center model for a laser.

\section*{Acknowledgments}


Work supported by the U.S. Department of Energy under SciDAC FNAL subcontract 644405, DE-SC0010064, and contract number DE-AC02-76SF00515, and by the U.S. National Science Foundation under NSF 1806046. The simulations were performed on the UCLA Hoffman 2 and Dawson 2 Clusters, the resources of the National Energy Research Scientific Computing Center, and the Super Computing Center of Beijing Normal University. 

\begin{appendix}
\section{Implementation of free boundary conditions for electromagnetic fields}
\label{sec:app:free_bnd}

\begin{figure}[htbp]
\centering
\includegraphics[width=0.6\textwidth]{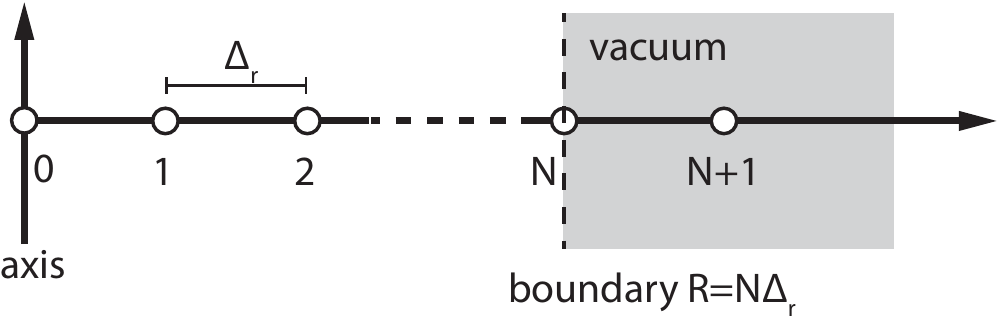}
\caption{Grid points layout in $r$-direction.}
\label{fig:boundary}
\end{figure}

In this appendix, we describe the implementation of the open (free) boundary conditions used to solve Eqs. (\ref{eq:pois_psi_m})-(\ref{eq:pois_bz_m}) and (\ref{eq:pois_ez_m}). Figure \ref{fig:boundary} shows the grid setup for solving the fields with total $N$ grid points within the solution region. The dashed line defines the boundary and the the physical domain. It is assume that outside of this region there is vacuum out to infinity.

The basic idea is to obtain the analytic solution in vacuum by solving Laplace equations and then applying solutions at the boundary. We first consider the scalar Laplace equation,
\begin{equation}
\label{eq:laplace_scalar}
\Laplace_m U^m=0,\quad\text{for}\ r>R,
\end{equation}
where $U^m$ represents the $m$th mode of $\psi$, $A_z$, $B_z$ and $E_z$. It has the solution
\begin{equation}
\label{eq:laplace_sol_0}
U^0=C_{U,0}+D_{U,0}\ln(r)
\end{equation}
and
\begin{equation}
U^{m>0}=C_{U,m} r^{-m}+D_{U,m} r^m.
\end{equation}
The determination of the constants $C_{U,m}$ and $D_{U,m}$ differs depending on the types of fields. For $\psi$, it can be shown that $D_{\psi,0}=0$. By applying Gauss's theorem to Eq. (\ref{eq:poisson_psi}) and considering a circular  region $S$ of integration with a radius greater than $R$, leads to
\begin{equation}
\label{eq:gauss_theorem}
\oint_{\partial S} \nabla_\perp\psi\ dl=2\pi r\PP{\psi^0}{r}=-\int_S (\rho-J_z) dS.
\end{equation}
Note that the $m>0$ modes of $\psi$ do not contribute to the integral on the left because of the presence of the term $e^{\ii m\phi}$. From the continuity equation under the QSA
\begin{equation}
\PP{}{\xi}(\rho-J_z)+\nabla_\perp\cdot\bm{J}_\perp=0.
\end{equation}
and using the fact that $\bm{J}_\perp$ vanishes at the boundary of the surface integral $\partial S$, we have
\begin{equation}
\PP{}{\xi}\int_S (\rho-J_z)dS=0.
\end{equation}
which indicates this integral is zero for any $\xi$ because it is initially zero (neutral plasma). Therefore, according to Eq. (\ref{eq:gauss_theorem}), we have $2\pi r\PP{\psi^0}{r}|_{r>R}=0$ which gives $D_{\psi,0}=0$ by inserting Eq. (\ref{eq:laplace_sol_0}). Requiring $\psi\rightarrow 0$ while $r\rightarrow 0$, we can determine that $C_{\psi,0}=0$ and $D_{\psi,m}=0\ (m>0)$, and thus the solution in the vacuum has the form
\begin{equation}
\psi^0=0,\quad
\psi^m=\frac{C_{\psi,m}}{r^m}.
\end{equation}

For the longitudinal component of beam's vector potential $A_z$, $D_{A_z,0}\neq0$ because the charge of the beam is apparently non-neutralized. Applying the natural boundary condition $B_{\phi,\text{beam}}=\PP{A_z}{r}\rightarrow0$ when $r\rightarrow0$ and ignoring the arbitrary constant, we have
\begin{equation}
A_z^0=D_{A_z,0}\ln(r),\quad
A_z^m=\frac{C_{A_z,m}}{r^m}.
\end{equation}

For $B_z$ and $E_z$, the only constraint is $B_z,E_z\rightarrow 0$ when $r\rightarrow 0$, so that
\begin{equation}
E_z^0=0,\quad
E_z^m=\frac{C_{E_z,m}}{r^m}
\end{equation}
and
\begin{equation}
B_z^0=0,\quad
B_z^m=\frac{C_{B_z,m}}{r^m}.
\end{equation}

The transverse magnetic fields induced by the plasma satisfy the coupled Laplace equations in the vacuum,
\begin{align}
\Laplace_m B_r^m-\frac{B_r^m}{r^2}-\frac{2\ii m}{r^2}B_\phi^m&=0, \\
\Laplace_m B_\phi^m-\frac{B_\phi^m}{r^2}+\frac{2\ii m}{r^2}B_r^m&=0.
\end{align}
It can be verified that the general solution can be written as
\begin{equation}
B_r^0=\frac{C_{B_r,0}}{r},\ B_r^m=\frac{C_{B_r,m}}{r^{m+1}}+D_{B_r,m}r^{m-1},
\end{equation}
and
\begin{equation}
B_\phi^0=\frac{C_{B_\phi,0}}{r},\ B_\phi^m=\frac{C_{B_\phi,m}}{r^{m+1}}+D_{B_\phi,m}r^{m-1}.
\end{equation}
Here, $D_{B_r,m}=D_{B_\phi,m}=0$ because of the natural boundary conditions that $B_r,B_\phi\rightarrow 0$ when $r\rightarrow 0$.

After obtaining the analytical solution for each field (components) in the vacuum, we derive the finite difference form of the boundary conditions used for solving the discrete Poisson-like equations. For an arbitrary field $U^m$, the value on the ghost cell can be evaluated through Taylor expansion (central difference)
\begin{equation}
U^m_{N+1}=U^m_N+\left.\PP{U^m}{r}\right|_N\Delta_r+O(\Delta_r^2).
\end{equation}
The derivative at $r_{N+\hf}$ (note that $r_{N+\hf}=R$) is evaluated using the analytical formula. For $A_z^0$,
\begin{equation}
A^0_{z,N+1}\simeq A^0_{z,N}+\frac{D_{A_z,0}}{R}\Delta_r\simeq A_{z,N}^0+\frac{\Delta_r}{R\ln R}A^0_{z,N+1}
\end{equation}
therefore
\begin{equation}
A^0_{z,N+1}\simeq\left(1+\frac{\Delta_r}{R\ln(R)}\right)A^0_{z,N}.
\end{equation}
Similarly, for $m>0$ modes of $\psi,A_z,B_z$ and $E_z$, we can obtain
\begin{equation}
\begin{pmatrix}
\psi \\ A_z \\ B_z \\ E_z
\end{pmatrix}^m_{N+1}\simeq\left(1-\frac{m\Delta_r}{R}\right)
\begin{pmatrix}
\psi \\ A_z \\ B_z \\ E_z
\end{pmatrix}^m_N
\end{equation}
and for all the modes of $B_r$ and $B_\phi$ associated with plasma
\begin{equation}
\begin{pmatrix}
B_r \\ B_\phi
\end{pmatrix}^m_{N+1}\simeq\left(1-\frac{(m+1)\Delta_r}{R}\right)
\begin{pmatrix}
B_r \\ B_\phi
\end{pmatrix}^m_N.
\end{equation}
As $B_\pm^m$ rather that $B_r^m$ and $B_\phi^m$ are directly solved in QPAD, we need to perform the linear transformation  $B_\pm^m=B_r^m\pm\ii B_\phi^m$ on both sides of the above equation to obtain  the boundary condition for $B_\pm^m$
\begin{equation}
B_{\pm,N}^m\simeq\left(1-\frac{(m+1)\Delta_r}{R}\right)B_{\pm,N+1}^m.
\end{equation}

\end{appendix}

\bibliographystyle{elsarticle-num}
\bibliography{refs}

\begin{thebibliography}{10}
\expandafter\ifx\csname url\endcsname\relax
  \def\url#1{\texttt{#1}}\fi
\expandafter\ifx\csname urlprefix\endcsname\relax\def\urlprefix{URL }\fi
\expandafter\ifx\csname href\endcsname\relax
  \def\href#1#2{#2} \def\path#1{#1}\fi

\bibitem{blumenfeld2007}
I.~Blumenfeld, et~al., Energy doubling of 42 gev electrons in a metre-scale
  plasma wakefield accelerator, Nature 445~(7129) (2007) 741--744.

\bibitem{litos2014}
M.~Litos, et~al., High-efficiency acceleration of an electron beam in a plasma
  wakefield accelerator, Nature 515~(7525) (2014) 92--95.

\bibitem{faure2004}
J.~Faure, et~al., A laser-plasma accelerator producing monoenergetic electron
  beams, Nature 431~(7008) (2004) 541--544.

\bibitem{geddes2004}
C.~G.~R. Geddes, et~al., High-quality electron beams from a laser wakefield
  accelerator using plasma-channel guiding, Nature 431~(7008) (2004) 538--541.

\bibitem{mangles2004}
S.~P.~D. Mangles, et~al., Monoenergetic beams of relativistic electrons from
  intense laser-plasma interactions, Nature 431~(7008) (2004) 535--538.

\bibitem{gonsalves2011}
A.~J. Gonsalves, et~al., Tunable laser plasma accelerator based on longitudinal
  density tailoring, Nature Physics 7~(11) (2011) 862--866.

\bibitem{wang2013}
X.~Wang, et~al., Quasi-monoenergetic laser-plasma acceleration of electrons to
  2 gev, Nature Communications 4~(1) (2013) 1988--1988.

\bibitem{steinke2016}
S.~Steinke, et~al., Multistage coupling of independent laser-plasma
  accelerators, Nature 530~(7589) (2016) 190--193.

\bibitem{guenot2017}
D.~Guénot, et~al., Relativistic electron beams driven by k{H}z single-cycle
  light pulses, Nature Photonics 11~(5) (2017) 293--296.

\bibitem{adli2018}
E.~Adli, et~al., Acceleration of electrons in the plasma wakefield of a proton
  bunch, Nature 561~(7723) (2018) 363.

\bibitem{corde2015}
S.~Corde, et~al., Multi-gigaelectronvolt acceleration of positrons in a
  self-loaded plasma wakefield, Nature 524~(7566) (2015) 442--445.

\bibitem{kneip2010}
S.~Kneip, et~al., Bright spatially coherent synchrotron x-rays from a table-top
  source, Nature Physics 6~(12) (2010) 980--983.

\bibitem{cipiccia2011}
S.~Cipiccia, et~al., Gamma-rays from harmonically resonant betatron
  oscillations in a plasma wake, Nature Physics advance online publication.

\bibitem{nie2018}
Z.~Nie, et~al., Relativistic single-cycle tunable infrared pulses generated
  from a tailored plasma density structure, Nature Photonics 12~(8) (2018)
  489--494.

\bibitem{dawson1983}
J.~M. Dawson, Particle simulation of plasmas, Reviews of Modern Physics 55~(2)
  (1983) 403--447.

\bibitem{birdsall2005}
C.~K. Birdsall, A.~B. Langdon, Plasma Physics via Computer Simulation,
  Institute of Physics Pub., 2005.

\bibitem{hockney1988}
R.~W. Hockney, J.~W. Eastwood, Computer simulation using particles, crc Press,
  1988.

\bibitem{decker1994}
C.~D. Decker, W.~B. Mori, Group velocity of large amplitude electromagnetic
  waves in a plasma, Physical Review Letters 72 (1994) 490--493.

\bibitem{top500}
\href{https://www.top500.org/lists/2018/11/}{Top 500 supercomputer sites}.
\newline\urlprefix\url{https://www.top500.org/lists/2018/11/}

\bibitem{xu2013}
X.~Xu, et~al., Numerical instability due to relativistic plasma drift in em-pic
  simulations, Computer Physics Communications 184~(11) (2013) 2503--2514.

\bibitem{xu2019}
X.~Xu, et~al., On numerical errors to the fields surrounding a relativistically
  moving particle in pic codes, arXiv preprint arXiv:1910.13529.

\bibitem{rosenzweig2005}
J.~B. Rosenzweig, et~al., Effects of ion motion in intense beam-driven plasma
  wakefield accelerators, Physical Review Letters 95~(19) (2005) 195002.

\bibitem{gholizadeh2010}
R.~Gholizadeh, et~al., Preservation of beam emittance in the presence of ion
  motion in future high-energy plasma-wakefield-based colliders, Physical
  Review Letters 104~(15) (2010) 155001.

\bibitem{an2017}
W.~An, et~al., Ion motion induced emittance growth of matched electron beams in
  plasma wakefields, Physical Review Letters 118~(24) (2017) 244801.

\bibitem{vay2007}
J.~L. Vay, Noninvariance of space- and time-scale ranges under a lorentz
  transformation and the implications for the study of relativistic
  interactions, Physical Review Letters 98~(13) (2007) 130405.

\bibitem{mora1997}
P.~Mora, J.~Thomas M.~Antonsen, Kinetic modeling of intense, short laser pulses
  propagating in tenuous plasmas, Physics of Plasmas 4~(1) (1997) 217--229.

\bibitem{whittum1997}
D.~H. Whittum, Transverse two-stream instability of a beam with a bennett
  profile, Physics of Plasmas 4~(4) (1997) 1154--1159.

\bibitem{lotov2003}
K.~V. Lotov, Fine wakefield structure in the blowout regime of plasma wakefield
  accelerators, Physical Review Special Topics - Accelerators and Beams 6~(6)
  (2003) 061301.

\bibitem{huang2006}
C.~Huang, et~al., {QUICKPIC}: A highly efficient particle-in-cell code for
  modeling wakefield acceleration in plasmas, Journal of Computational Physics
  217~(2) (2006) 658--679.

\bibitem{an2013}
W.~An, et~al., An improved iteration loop for the three dimensional
  quasi-static particle-in-cell algorithm: Quick{PIC}, Journal of Computational
  Physics 250 (2013) 165--177.

\bibitem{mehrling2014}
T.~Mehrling, et~al., Hi{PACE}: a quasi-static particle-in-cell code, Plasma
  Physics and Controlled Fusion 56~(8) (2014) 084012.

\bibitem{lifschitz2009}
A.~F. Lifschitz, et~al., Particle-in-cell modeling of laser–plasma
  interaction using fourier decomposition, Journal of Computational Physics
  228~(5) (2009) 1803--1814.

\bibitem{davidson2015}
A.~Davidson, et~al., Implementation of a hybrid particle code with a pic
  description in r–z and a gridless description in $\phi$ into {OSIRIS},
  Journal of Computational Physics 281 (2015) 1063--1077.

\bibitem{lehe2016}
R.~Lehe, et~al., A spectral, quasi-cylindrical and dispersion-free
  particle-in-cell algorithm, Computer Physics Communications 203 (2016)
  66--82.

\bibitem{yu2016}
P.~Yu, et~al., Enabling lorentz boosted frame particle-in-cell simulations of
  laser wakefield acceleration in quasi-3d geometry, Journal of Computational
  Physics 316 (2016) 747--759.

\bibitem{sprangle1990a}
P.~Sprangle, et~al., Nonlinear interaction of intense laser pulses in plasmas,
  Physical Review A 41 (1990) 4463--4469.

\bibitem{sprangle1990b}
P.~Sprangle, et~al., Nonlinear theory of intense laser-plasma interactions,
  Physical Review Letters 64 (1990) 2011--2014.

\bibitem{feng2009}
B.~Feng, et~al., Enhancing parallel quasi-static particle-in-cell simulations
  with a pipelining algorithm, Journal of Computational Physics 228~(15) (2009)
  5340--5348.

\bibitem{fonseca2002}
R.~A. Fonseca, et~al., {OSIRIS}: A three-dimensional, fully relativistic
  particle in cell code for modeling plasma based accelerators, in:
  International Conference on Computational Science, Springer, 2002, pp.
  342--351.

\bibitem{ferri2018}
J.~Ferri, et~al., High-brilliance betatron $\gamma$-ray source powered by
  laser-accelerated electrons, Physical Review Letters 120~(25) (2018) 254802.

\bibitem{press2010}
W.~H. Press, et~al., Numerical Recipes: The Art of Scientific Computing (3rd
  Edition), Cambridge University Press, 2010.

\bibitem{nielson1976}
C.~W. Nielson, H.~R. Lewis, Particle code models in the nonradiative limit,
  Methods in Computational Physics 16 (1976) 367.

\bibitem{boris1972}
J.~P. Boris, R.~A. Shanny, Proceedings: Fourth Conference on Numerical
  Simulation of Plasmas, November 2, 3, 1970, Naval Research Laboratory, 1972.

\bibitem{qin2013}
H.~Qin, et~al., Why is {Boris} algorithm so good?, Physics of Plasmas 20~(8)
  (2013) 084503.

\bibitem{falgout2002}
R.~D. Falgout, U.~M. Yang, hypre: A library of high performance
  preconditioners, in: International Conference on Computational Science,
  Springer, 2002, pp. 632--641.

\bibitem{constantinescu2002}
G.~S. Constantinescu, S.~K. Lele, A highly accurate technique for the treatment
  of flow equations at the polar axis in cylindrical coordinates using series
  expansions, Journal of Computational Physics 183~(1) (2002) 165--186.

\bibitem{lu2006}
W.~Lu, et~al., Nonlinear theory for relativistic plasma wakefields in the
  blowout regime, Physical Review Letters 96~(16) (2006) 165002.

\bibitem{rosenzweig1991}
J.~B. Rosenzweig, et~al., Acceleration and focusing of electrons in
  two-dimensional nonlinear plasma wake fields, Physical Review A 44~(10)
  (1991) R6189--R6192.

\bibitem{whittum1991}
D.~H. Whittum, et~al., Electron-hose instability in the ion-focused regime,
  Physical Review Letters 67~(8) (1991) 991.

\bibitem{huang2007}
C.~Huang, et~al., Hosing instability in the blow-out regime for
  plasma-wakefield acceleration, Physical review letters 99~(25) (2007) 255001.

\end{thebibliography}

\end{document}